\pdfoutput=1 

\documentclass[12pt,tightenlines,nofootinbib,showpacs,showkeys]{revtex4}

\usepackage{amsmath}
\usepackage{amsfonts}
\usepackage{bm}
\usepackage{euscript}
\usepackage{wasysym}
\usepackage{graphicx}
\usepackage{subfigure}
\usepackage{color}
\usepackage{lscape}
\usepackage{rotating}
\usepackage{framed}
\usepackage[english]{babel}

\newcommand{\qed}{\nobreak \ifvmode \relax \else
      \ifdim\lastskip<1.5em \hskip-\lastskip
      \hskip1.5em plus0em minus0.5em \fi \nobreak
      \vrule height0.75em width0.5em depth0.25em\fi}

\newcommand{\mathd}{\mathrm{d}}
\newcommand{\mathe}{\mathrm{e}}

\newcommand{\bd}{\bm{\mathcal{B}}}

\newcommand{\angleS}{\varphi}

\newcommand{\ampl}{A_0}

\newcommand{\kcorr}{c}

\newcommand{\pot}{\mathcal{U}}

\newcommand{\xs}{X}
\newcommand{\ys}{Y}
\newcommand{\zs}{Z}

\begin{document}
%+Title
\title{Modelling the growth rate of a tracer gradient using stochastic differential equations}
\author{Lennon \'O N\'araigh$\,^{1}$\footnote{Corresponding author.   Tel.: +353 1 716 2546; Fax: +353 1 716 1172.\\Email address: lennon.onaraigh@ucd.ie}}
\affiliation{
School of Mathematical Sciences, University College Dublin, Belfield, Dublin 4, Ireland
}
\date{\today}
%-Title
%
%

\begin{abstract}
We develop a model in two dimensions to characterise the growth rate of a tracer gradient mixed by a statistically homogeneous flow with rapid temporal variations.
The model is based on the orientation dynamics of the passive-tracer gradient with respect to the straining (compressive) direction of the flow, and involves reducing the dynamics to a set of stochastic differential equations. The statistical properties of the system emerge from solving the associated Fokker--Planck equation.   In a certain limiting case, and within the model framework, there is a rigorous proof that the tracer gradient aligns with the compressive direction.  This limit involves decorrelated flows whose mean vorticity is zero.
Using numerical simulations, we assess the extent to which our model applies to real mixing protocols, and map the stochastic parameters on to flow parameters.
\end{abstract}


\maketitle
\section{Introduction}

\noindent 
When a passive tracer is stirred by a flow that varies rapidly in space and time, stochastic models are often used to describe the mixing~\cite{Falkovich2001}.  Based on the Fokker--Planck (FP) equation, we develop a model for the probability distribution function (PDF) of the growth rate of the tracer gradient in a generic, two-dimensional, rapidly-varying stirring flow.  
%
% The use of the FP equation is computationally cheap and highlights the striking analogies between the current work and % a diversity of other physical systems~\cite{RiskenBook}.
%
In this Introduction, we place our model in the context of existing literature.  

%\textit{Stochastic models of turbulence:}  
%
Kraichnan~\cite{Kraichnan1974} modelled homogeneous isotropic turbulence as Gaussian fluctuations.  In this description, the PDF for the largest finite-time Lyapunov exponent (FTLE) can be found  using the Central Limit Theorem.   The same model has been used to describe the so-called strange eigenmode~\cite{Pierrehumbert1994} in passive-tracer decay~\cite{Schekochihin2004}.
Balkovsky and Fouxon~\cite{Balkovsky1999} generalised the Kraichnan model to account for temporally-correlated flows.  When the correlation time tends to zero, their analysis yields the PDF of all the FTLEs (in $D$ dimensions)~\cite{Balkovsky1999,Bernard1998}.   For a review of these techniques, see~\cite{Falkovich2001}.  In the present work, we focus instead on the distribution of the growth rate of the tracer gradient, a distinct but related quantity.

% \textit{Orientation dynamics:}  
%
Lapeyre \textit{et al.}~\cite{Lapeyre1999} derived a pair of equations to describe the growth rate of the gradient of a passively-advected tracer.  The theory is based on the orientation of the tracer gradient relative to the straining (compressive) direction of the flow: if the tracer gradient aligns with the compressive direction, it blows up.  However, the tracer-gradient orientation is instead fixed by a balance between strain and `effective' vorticity (the sum of vorticity and twice the rotation rate of the strain eigenbasis).  Thus, blowup is prevented in rotation-dominated regions. 
The inclusion of the rotation rate generalises the description of orientation dynamics implied by the  Okubu--Weiss criterion~\cite{Weiss1991}.  That said, the Lapeyre criterion relies on the adiabatic approximation of the differential equations: the effective vorticity and strain eigenvalue are assumed to vary slowly along trajectories.  Although this description is valid in certain cases (e.g.~\cite{Lapeyre2002}), it may not hold when the stirring flow is governed by some kind of externally-imposed, rapidly-varying forcing: this will be our focus herein.

Gonzalez and Parantho\"en~\cite{Gonzalez2010} studied two mixing protocols wherein the adiabatic assumption breaks down.  They show numerically that the tracer gradient points a direction fixed by the average value of the ratio $r$ of effective vorticity to strain.  
Garcia and coworkers~\cite{Garcia2005} introduced a stochastic model of such scenarios by regarding $r$ as the solution to a stochastic differential equation (SDE).  They found good agreement with an experiment they performed.  However, they neither examined the distribution of growth rates of the tracer gradient elicited by this model (but see~\cite{Gonzalez2009}), nor analysed the associated FP equation.
In the present work, we address these issues and provide a more detailed model of a generic flow  (Sec.~\ref{sec:model_new}).
We make the connection with the FP equation~\cite{RiskenBook}, an approach that facilitates fast, accurate numerical calculations, analytical results in a certain limiting case, and highlights the striking analogies between the current work and a variety of other physical systems~\cite{RiskenBook}.
Furthermore, our approach involves a comparison between theoretical models and simulations of the random-phase sine flow (Sec.~\ref{sec:sineflow}), and forced, two-dimensional turbulence (Sec.~\ref{sec:turb}), an exercise that strengthens the case for model developed herein.
First, we review the orientation dynamics derived elsewhere~\cite{Lapeyre1999,Dresselhaus1992}.  

\section{The orientation dynamics}
\label{sec:model_old}

The description of the orientation dynamics starts with the vector field $\bd=\left(-\theta_y,\theta_x\right)$, where $\theta$ is the passively-advected tracer
\begin{equation}
\theta_t+\bm{u}\cdot\nabla\theta=\kappa\Delta\theta,\qquad \nabla\cdot\bm{u}=0.
\label{eq:scalar}
\end{equation}
In this section, we take the molecular diffusivity $\kappa$ to be zero;  we study the implications of finite $\kappa$-values at the end of the derivation.
When $\kappa=0$, the vector $\bd$ satisfies the material-line equation
\begin{equation}
\bd_t+\bm{u}\cdot\nabla\bd=\bd\cdot\nabla\bm{u}.
\label{eq:bd1}
\end{equation}
This result can be obtained by brute-force calculation based on Eq.~\eqref{eq:scalar}, or by other means~\cite{Holm2010}
\footnote{Indeed, one may regard $\bd$ as a complex-valued two-form $\mathcal{B}=\mathcal{B}_1+\mathrm{i}\mathcal{B}_2$ and operate on both sides of the two-form identity $\mathcal{B}\mathd z\wedge\mathd\overline{z}=2\mathrm{i}\mathd{z}\wedge\mathd\theta$ with the material derivative.}.  
%
%
%Dotting Eq.~\eqref{eq:bd1} with $\bd$, we obtain
%
%
%
%
\begin{equation}
\frac{\mathd}{\mathd t}|\bd|^2=\left(\mathcal{B}_1,\mathcal{B}_2\right)\left(\begin{array}{cc}s&d\\d&-s\end{array}\right)\left(\begin{array}{c}\mathcal{B}_1\\\mathcal{B}_2\end{array}\right).
\label{eq:bd2}
\end{equation}
The rate-of-strain matrix $S$ appears in Eq.~\eqref{eq:bd2}: $s=u_x$ and $d=\left(u_y+v_x\right)/2$.
We identify $\beta-\left(\pi/2\right)$ as the phase of the vector $\bd$, where $\tan\beta=\theta_y/\theta_x$, and obtain, by means similar to before, an equation for $\beta$:
\begin{equation}
\frac{\mathd\beta}{\mathd t}=\tfrac{1}{2}\omega-\frac{1}{|\bd|^2}\left(\mathcal{B}_2,-\mathcal{B}_1\right)\left(\begin{array}{cc}s&d\\d&-s\end{array}\right)\left(\begin{array}{c}\mathcal{B}_1\\\mathcal{B}_2\end{array}\right),
\label{eq:beta}
\end{equation}
where $\omega=v_x-u_y$ is the vorticity.
To complete our analysis, we re-write Eq.~\eqref{eq:bd2} in terms of the eigenvalues of $S$:
\begin{equation}
\frac{\mathd}{\mathd t}|\bd|^2=-2\lambda\sin\zeta\,|\bd|^2.
\label{eq:bd_eigen}
\end{equation}
Here, $\lambda=\text{sign}(d)\sqrt{s^2+d^2}$ is the \textit{unsigned} eigenvalue of $S$, with associated eigenvector $\bm{X}_{(+)}$.  The angle between the $x$-axis and $\bm{X}_{(+)}$ is $\varphi$, 
and $\zeta=2(\beta-\angleS+\tfrac{1}{4}\pi)$.
Hence, from Eq.~\eqref{eq:beta}
\begin{equation}
\frac{\mathd\zeta}{\mathd t}=-2\lambda\cos\zeta+\omega+2\frac{\mathd\varphi}{\mathd t}.
\label{eq:angle}
\end{equation}
Finally, we identify the growth rate of the gradient:
\begin{equation}
\Lambda:=-2\lambda\sin\zeta.
\end{equation}
To understand this description, imagine a situation where $\lambda>0$.  Then the growth rate is maximised when $\zeta=-\pi/2$, or $\beta=\angleS-\pi/2$, that is, when the tracer gradient aligns with the \textit{compressive} direction of the flow.  In this paper, our concern is the extent to which this alignment condition is fulfilled.

The basic equations~\eqref{eq:bd2} and~\eqref{eq:angle} rely only on the assumptions $\kappa=0$ and incompressibility.  If $\kappa$ is finite but small, then the main effect of diffusion will be to reduce the growth rate of the gradient, through the presence of a negative term in Eq.~\eqref{eq:bd_eigen}.  We do not expect it to interfere much with the orientation dynamics~\cite{Lapeyre1999,Garcia2005}.  Hence, we expect similar results for the PDF of $-2\lambda\sin\zeta$, with or without a small amount of diffusion.  This is confirmed in our numerical simulations in Sec.~\ref{sec:turb}.  

\section{A stochastic model}
\label{sec:model_new}

\noindent For notational reasons, we write $X:=\zeta$~\cite{RiskenBook} and re-write Eq.~\eqref{eq:angle}:
\begin{equation}
\tfrac{1}{2}\frac{d\xs}{dt}=\left(-\lambda_0\cos \xs+w\right)-\ys\left(t\right)\cos \xs+\zs_0\left(t\right),
\label{eq:sde0}
\end{equation}
where $\lambda=\lambda_0+\ys\left(t\right)$ and $\left(\omega/2\right)+\dot\angleS=w+Z_0\left(t\right)$ represent decompositions into mean components and fluctuations.
We assume that the vector $\bm{X}_{(+)}$ spends an equal amount of time representing the compressive and expansive directions, hence $\lambda_0=0$.
We model the fluctuations as Ornstein--Uhlenbeck (OU) processes, with mean zero, common decay time $\tau$, and strengths $D_\ys$ and $D_\zs$ respectively (the fluctuations reduce to Wiener processes in the limit where the decay time tends to zero).  We anticipate this model to be appropriate in particular when the flow is governed by some random, externally-prescribed stirring mechanism (for examples, see~\cite{Molenaar2004}).
For generality, we assume that the noise terms in Eq.~\eqref{eq:sde0} satisfy the following correlation relations: 
\begin{eqnarray*}
\langle \ys\left(t\right)\ys\left(t'\right)\rangle&=&\frac{D_\ys}{\tau}e^{-|t-t'|/\tau},\\
\langle \zs_0\left(t\right)\zs_0\left(t'\right)\rangle&=&\frac{D_{\zs}}{\tau} e^{-|t-t'|/\tau},\\
\langle \ys\left(t\right)\zs_0\left(t'\right)\rangle&=&\kcorr\frac{\sqrt{D_\ys D_\zs}}{\tau}e^{-|t-t'|/\tau},
\end{eqnarray*}
where $0\leq \kcorr\leq 1$ is a correlation coefficient, and where $\tau^{-1}e^{-|s|/\tau}$ converges in the sense of distributions to $2\delta\left(s\right)$ as $\tau\rightarrow 0$.  Finally, we make the assumption that the underlying, noise-generating flow is homogeneous in space in a statistical sense:
\[
\langle \ys\left(t;\bm{x}_0\right)\ys\left(t';\bm{x}_0'\right)\rangle=\tau^{-1}\mathcal{D}_\ys\left(\bm{x}_0-\bm{x}_0'\right)e^{-|t-t'|/\tau},\qquad\text{\&c.}
\]
Hence,
%
%
%\begin{multline*}
\begin{equation*}
\langle \ys\left(t;\bm{x}_0\right)\ys\left(t';\bm{x}_0\right)\rangle=\tau^{-1}\mathcal{D}_\ys\left(0\right)e^{-|t-t'|/\tau}\\
:=\tau^{-1}D_\ys e^{-|t-t'|/\tau},\qquad\text{\&c.},
\end{equation*}
%\end{multline*}
%
%
and the noise strength is independent of the initial position of the Lagrangian particle.

Equation~\eqref{eq:sde0} is Markovian with uncorrelated noise terms if viewed in an augmented state space:
\begin{eqnarray}
\tfrac{1}{2}\frac{d\xs}{dt}&=&w+\left(-\cos\xs+\kcorr\delta^{1/2}\right)\ys+\zs,\nonumber\\
\frac{d\ys}{dt}&=&-\frac{\ys}{\tau}+\frac{\sqrt{D_\ys}}{\tau} \xi_\ys,\nonumber\\
\frac{d\zs}{dt}&=&-\frac{\zs}{\tau}+\frac{\sqrt{D_\zs\left(1-\kcorr^2\right)}}{\tau} \xi_\zs,
\label{eq:model_ou}
\end{eqnarray}
where $\xi_\ys$ and $\xi_\zs$ are uncorrelated Wiener processes of strength 2, and $\delta=D_\zs/D_\ys$.  Thus, $\ys$ and $\zs$ are completely uncorrelated, and the triple $\left(\xs,\ys,\zs\right)$ follows a Markov process, with an associated FP equation.
The form of Eqs.~\eqref{eq:model_ou} is similar to models used to describe pendulum motion in the overdamped limit, the orientation of electric dipoles in an external field, the Josephson junction~\cite{Li1998,BaroneBook}, and semiclassical lasers~\cite{RiskenBook}.  Note also that Eq.~\eqref{eq:model_ou} contains limiting cases that depend on the value of $\tau$.  For $\tau\rightarrow\infty$, we obtain the adiabatic regime described by Lapeyre~\cite{Lapeyre1999}.  Instead, the focus of this paper is on rapidly-forced regimes, wherein $\tau$ is either comparable to the root-mean-square values of the fluctuations, or tends to zero; the latter case gives rise to some analytical results concerning the orientation dynamics.

We compute the PDF of angles $\xs$, and growth rates $\Lambda$.  These 
are marginal PDFs that can be derived from the PDF $P\left(\xs,\ys,\zs,t\right)$, which satisfies the following FP equation:
\begin{equation}
\frac{\partial P}{\partial t}=\mathcal{L}_{OU}P-\frac{\partial}{\partial \xs}\left(VP\right),
\label{eq:fp}
\end{equation}
where
%
%
%\begin{multline*}
\begin{equation}
\mathcal{L}_{OU}=\frac{1}{\tau}\frac{\partial}{\partial y}\left(\ys\,\circ\right)+\frac{D_\ys}{\tau^2}\frac{\partial^2}{\partial \ys^2}+\\
\frac{1}{\tau}\frac{\partial}{\partial\zs}\left(\zs\,\circ\right)+
\frac{D_\zs\left(1-\kcorr^2\right)}{\tau^2}\frac{\partial^2}{\partial\zs^2}
\end{equation}
%\end{multline*}
%
%
is the OU operator associated with the $\xs-\ys$ variables, and 
\[
V=2\left[w+\left(-\cos\xs+\kcorr\delta^{1/2}\right)\ys+\zs\right]
\]
is the drift velocity.
We solve the \textit{stationary} FP equation with $X\in\left[-\pi,\pi\right]$ and $Y,Z\in\left(-\infty,\infty\right)$.  The PDF of angles is obtained from the stationary solution $P\left(\xs,\ys,\zs\right)$:
\[
P_\xs\left(\xs\right)=\int_{-\infty}^{\infty}\mathd\ys\int_{-\infty}^{\infty}\mathd\zs\, P\left(\xs,\ys,\zs\right).
\]
Since $\xs$ and $\ys$ are OU processes, we also have
%
%\begin{multline}
\begin{equation}
P_{\ys\zs}\left(\ys,\zs\right)=
\int_{-\pi}^{\pi}\mathd\xs\,P\left(\xs,\ys,\zs\right)\\\propto
e^{-\ys^2\tau/(2D_\ys)}e^{-\zs^2\tau/\left[2D_\zs\left(1-\kcorr^2\right)\right]}.
\label{eq:marg_yz}
\end{equation}
%\end{multline}
%
%
Finally, the PDF of growth rates $\Lambda=-2\ys\sin\xs$ is computed through a coordinate transformation of the $\ys$-variable
\begin{equation}
P_\Lambda\left(\Lambda\right)=\int_{-\pi}^{\pi}P_{\xs\ys}\left(\xs,\frac{\Lambda}{-2\sin\xs}\right)\frac{1}{2|\sin\xs|}\mathd\xs,
\label{eq:marg_lambda}
\end{equation}
and this transformation is legitimate because the Jacobian diverges only where $P$ vanishes, and $P$ vanishes rapidly as $\ys\rightarrow\pm\infty$.
\begin{figure}
\begin{center}
\subfigure[]{\includegraphics[width=0.45\textwidth]{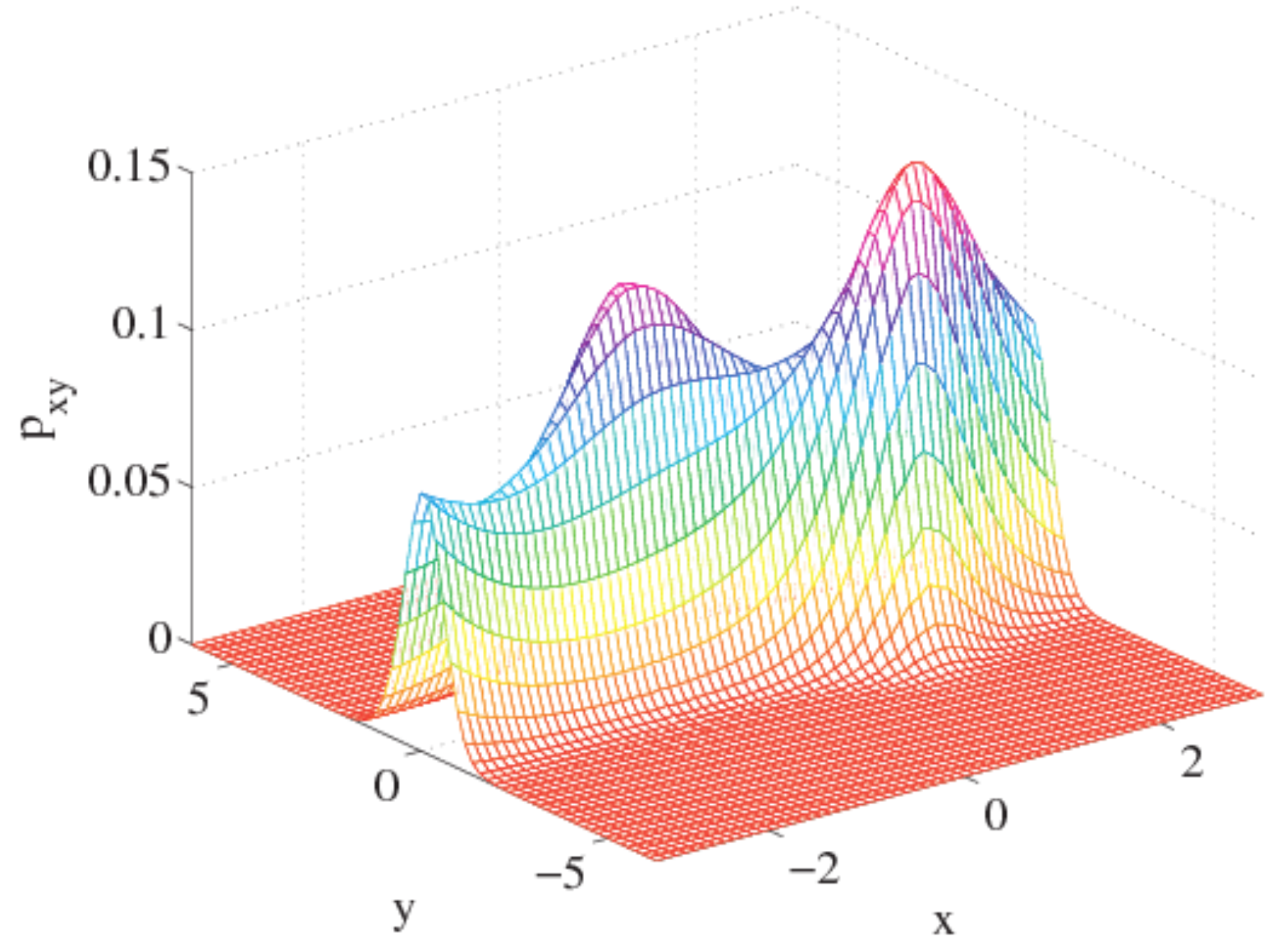}}
\subfigure[]{\includegraphics[width=0.45\textwidth]{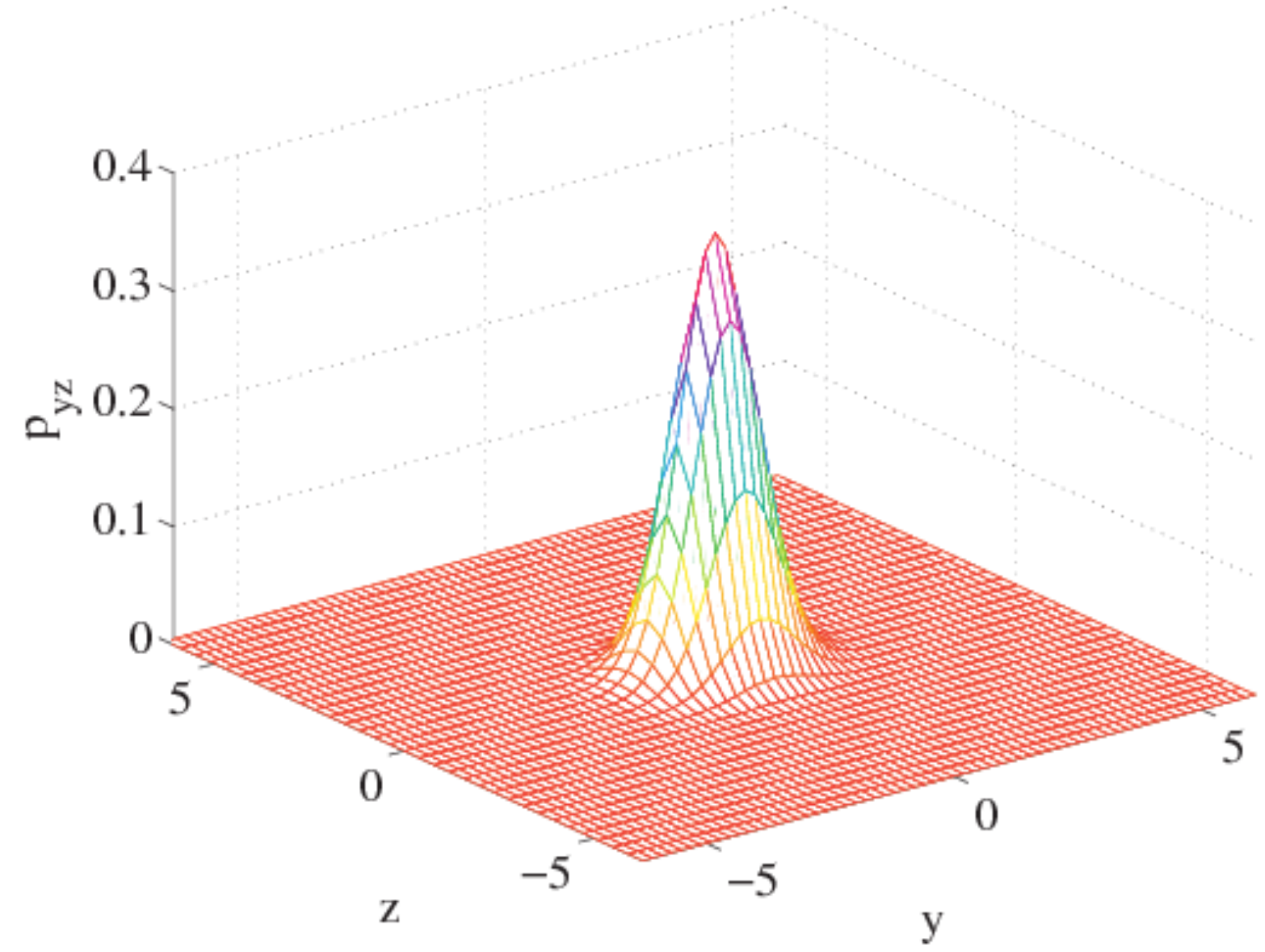}}
\end{center}
\caption{Marginal probability distributions obtained from the stationary FP equation.  (a) The distribution in $\xs$-$\ys$ space; (b) The distribution in $\ys$-$\zs$ space.  The latter is Gaussian, with decay scales in agreement with the theoretical values in Eq.~\eqref{eq:marg_yz}.}
\label{fig:marginal1}
\end{figure}

\begin{figure}
\begin{center}
\subfigure[]{\includegraphics[width=0.45\textwidth]{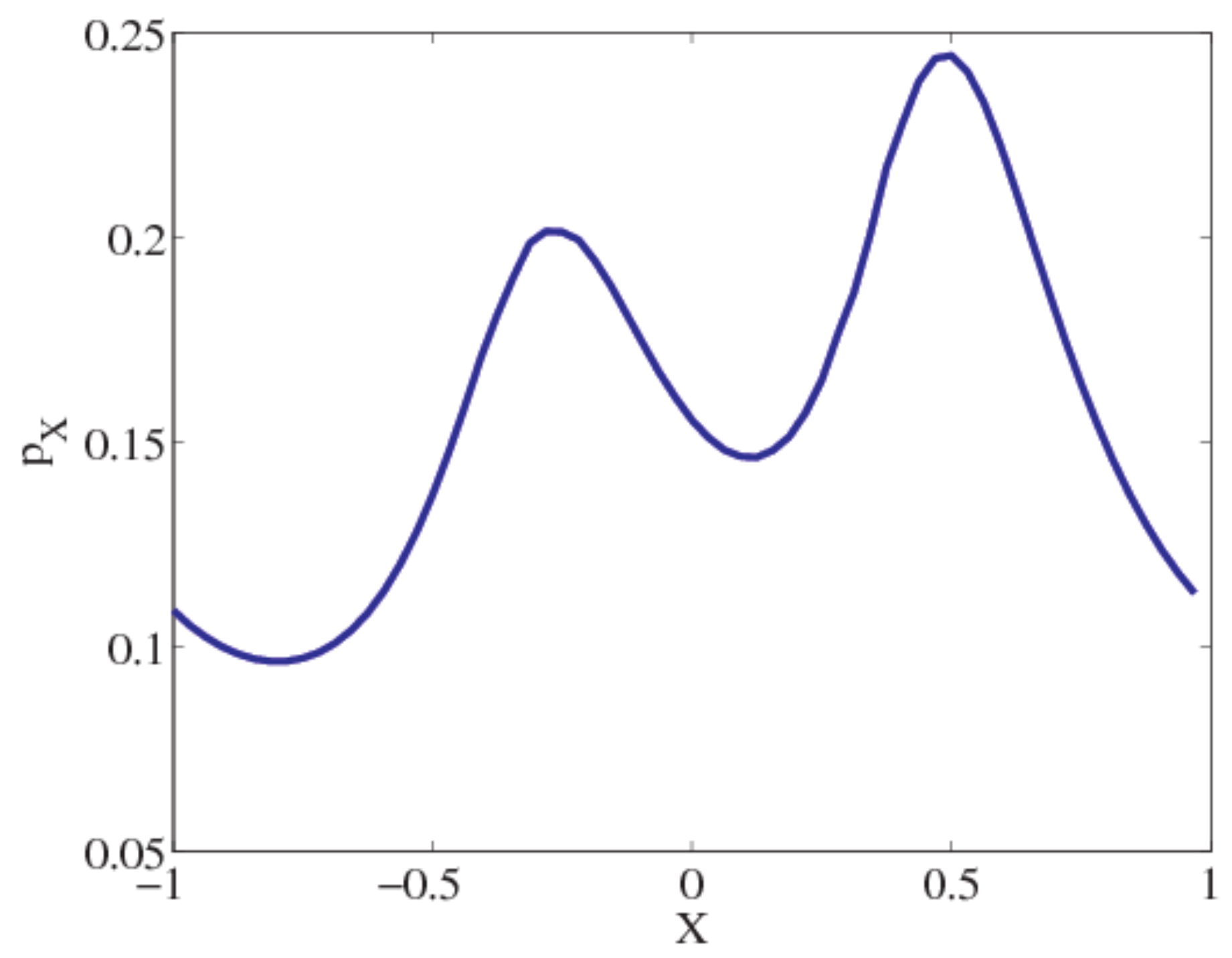}}
\subfigure[]{\includegraphics[width=0.45\textwidth]{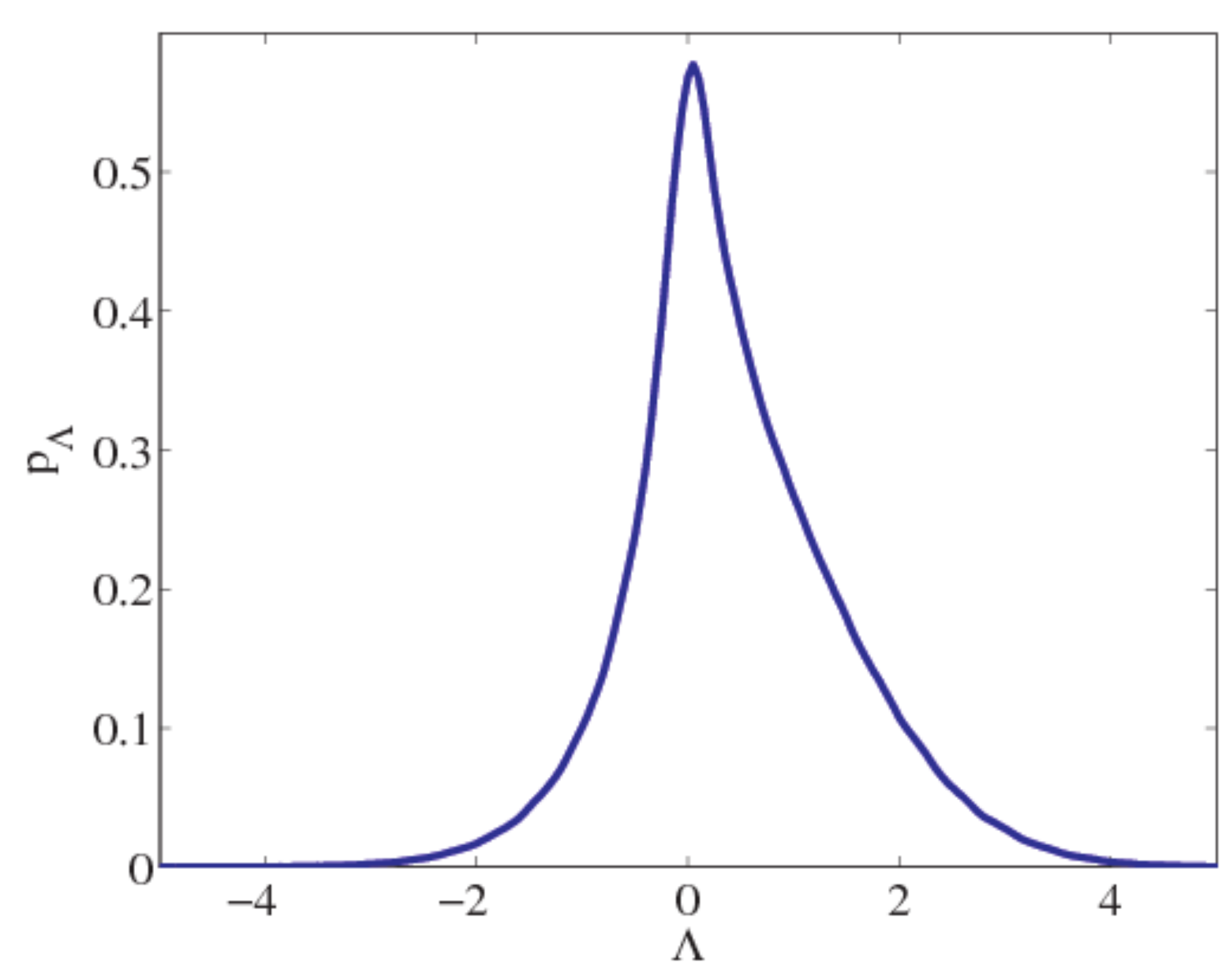}}
\end{center}
\caption{(a) The PDF of angles according to the SDE~\eqref{eq:sde0}, for nonzero values of $w$ and $\kcorr$.  The favoured orientations are shifted away from $\pm\pi/2$; (b) The PDF of growth rates.  A clear preference for positive growth is present, indicating that the tracer gradient prefers an orientation close to the compressive direction.  Here $w=\kcorr=0.5$ and $D_Y=D_Z=1$.
}
\label{fig:marginal2}
\end{figure}

Sample numerical results concerning the marginal PDFs are shown in Figs.~\ref{fig:marginal1}--\ref{fig:marginal2}.  The diffusion equation~\eqref{eq:fp} is solved using semi-implicit spectral method; as a validation of our technique, we compute the marginal PDF $P_{\ys\zs}$.  This is Gaussian, with decay scales in agreement with Eq.~\eqref{eq:marg_yz}.  Thus, we are satisfied with the accuracy of the method.  
The PDF of angles $\xs$ is computed from the full solution $P\left(\xs,\ys,\zs\right)$ using numerical integration: this possesses two maxima to the right of $\xs=\pm \pi/2$.  The shift in the maxima away from $\pm\pi/2$ is due to the finite value of $\kcorr$ in the calculation, while the non-invariance of the curve under $X\rightarrow-X$ is due to the finite value of the drift $w$.
When these parameters are set to zero, alignment with the directions $\pm\pi/2$ is achieved (Secs.~\ref{sec:sineflow}--\ref{sec:turb}).  Finally, the mean growth rate of the tracer gradient is positive, since the distribution of growth rates is skewed (Fig.~\ref{fig:marginal2}~(b)).  This suggests the favoured orientation is close to the compressive direction.
This situation corresponds to that described by Gonzalez and Parantho\"en~\cite{Gonzalez2010}, where in rapidly-varying flows, the tracer gradient aligns with a direction fixed by $\langle r\rangle$, which coincides with the compressive direction only when $\langle r\rangle=0$.

Lastly, we focus on the limiting case where $\tau\rightarrow 0$.  In this case, an explicit formula for the PDF of angles exists (up to quadratures).  This enables us to prove some rigorous results concerning the orientation of the tracer gradient relative to the compressive direction; it also facilitates a detailed parameter study to 
investigate whether the mean growth rate is always positive.
Now in the Wiener limit, the stationary PDF of angles satisfies the ordinary differential equation
\begin{equation}
J=A\left(\xs\right)P\left(\xs\right)-\frac{\mathd}{\mathd\xs}\left(P\left(\xs\right)B\left(\xs\right)\right),
\label{eq:jstat}
\end{equation}
where $J$ is the probability current and in the Stratonovich interpretation, the coefficients $A$ and $B$ have the following form:
\begin{eqnarray*}
A\left(\xs\right)&=&w+D_Yg'\left(\xs\right)\left[g\left(\xs\right)+\kcorr\delta^{1/2}\right],\\
B\left(\xs\right)&=&D_Z\left(1-\kcorr^2\right)+D_Y\left[g\left(\xs\right)+\kcorr\delta^{1/2}\right]^2,
\end{eqnarray*}
where $g\left(\xs\right)=-\cos\xs$.
Introducing the effective potential
\[
\pot\left(\xs\right)=-\int_{-\pi}^\xs\frac{A\left(\xs'\right)}{B\left(\xs'\right)}\mathd\xs',
\]
\[
P\left(\xs\right)=\frac{1}{B\left(\xs\right)}\mathe^{-\pot\left(\xs\right)}\left[N-J\int_a^\xs \mathe^{\pot\left(\xs'\right)}\mathd\xs'\right],
\]
where $N$ and $J$ are constants of integration fixed by normalisation and by the periodicity condition
\begin{equation}
J=\frac{B\left(-\pi\right)P\left(-\pi\right)\left[\mathe^{\pot\left(-\pi\right)}-\mathe^{\pot\left(\pi\right)}\right]}{\int_{-\pi}^{\pi}\mathe^{\pot\left(\xs\right)}\mathd\xs},
\label{eq:jperiodic}
\end{equation}
where $w\neq 0$ (the case $w=0$ is discussed below).
Thus, $P\left(\xs\right)$ is $2\pi$-periodic on $\left[-\pi,\pi\right]$~\cite{RiskenBook,Li1998}. 
An expression for the \textit{mean} growth rate $\langle\Lambda\rangle=-\langle 2\ys\sin\xs\rangle$ now follows from an application of the Forutsu--Novikov theorem~\cite{KonotopBook,Li1998}:
%
%
%
%
%\begin{multline*}
\begin{equation*}
\langle \ys\left(t\right)g^{(n)}\left(\xs\right)\rangle=4D_Y\int_{-\infty}^{\infty}\mathd{t}'
\delta\left(t-t'\right)\Big\langle g^{(n+1)}\left(\xs\left(t'\right)\right)\left[g\left(\xs\left(t'\right)\right)+\kcorr\delta^{1/2}\right]\Big\rangle.
\end{equation*}
%\end{multline*}
%
%
Setting $n=1$ gives and recalling that $\lambda_0=0$, we obtain the identity
\begin{equation}
\langle\Lambda\rangle={8D_Y}\left(\langle \cos^2 \xs\rangle-\kcorr\delta^{1/2}\langle\cos \xs\rangle\right).
\label{eq:lambda_sde}
\end{equation}
Two analytical results arise from this formalism.  First, \textit{when $c=0$, the mean growth rate is definitely non-negative}.  Moreover, if $w=c=0$, it follows that $J=0$.  Thus, $P_X\propto B\left(X\right)^{-1/2}$, and $P_X$ attains its maximum at $X=\pm\pi/2$.  In other words, \textit{for uncorrelated forcing with zero mean vorticity, the tracer aligns with the compressive direction, on average}.  In this case,
\begin{equation}
\langle \Lambda\rangle=8D_Y\delta\left[\left(1+\frac{1}{\delta}\right)\frac{E\left(\pi|\frac{1}{\delta+1}\right)}{F\left(\pi|\frac{1}{\delta+1}\right)}-1\right],
\end{equation}
where $E$ and $F$ are incomplete elliptic functions of the first and second kinds, respectively.

\begin{figure}
\begin{center}
\subfigure[$\,w=0$]{\includegraphics[width=0.32\textwidth]{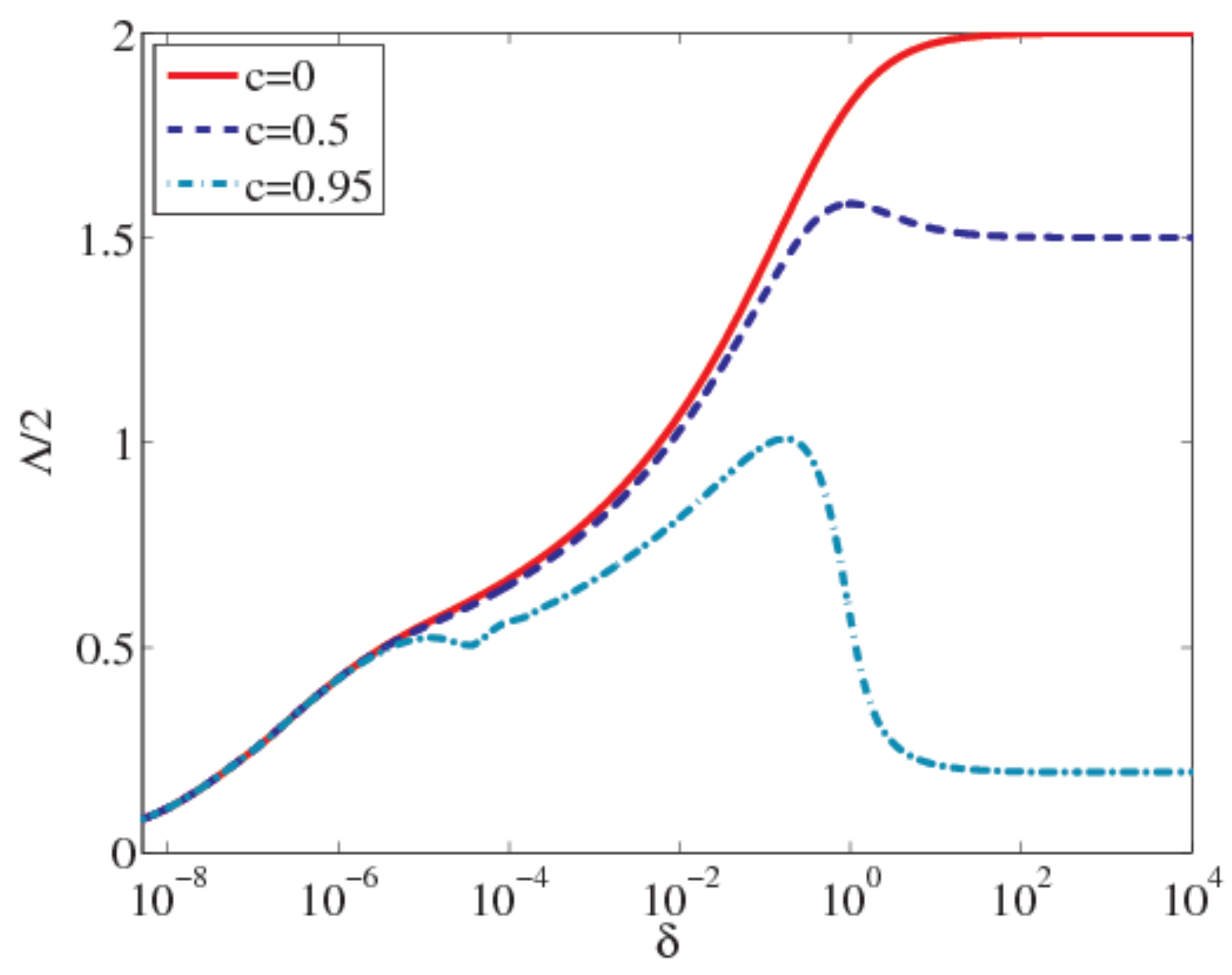}}
\subfigure[$\,w=0.5$]{\includegraphics[width=0.32\textwidth]{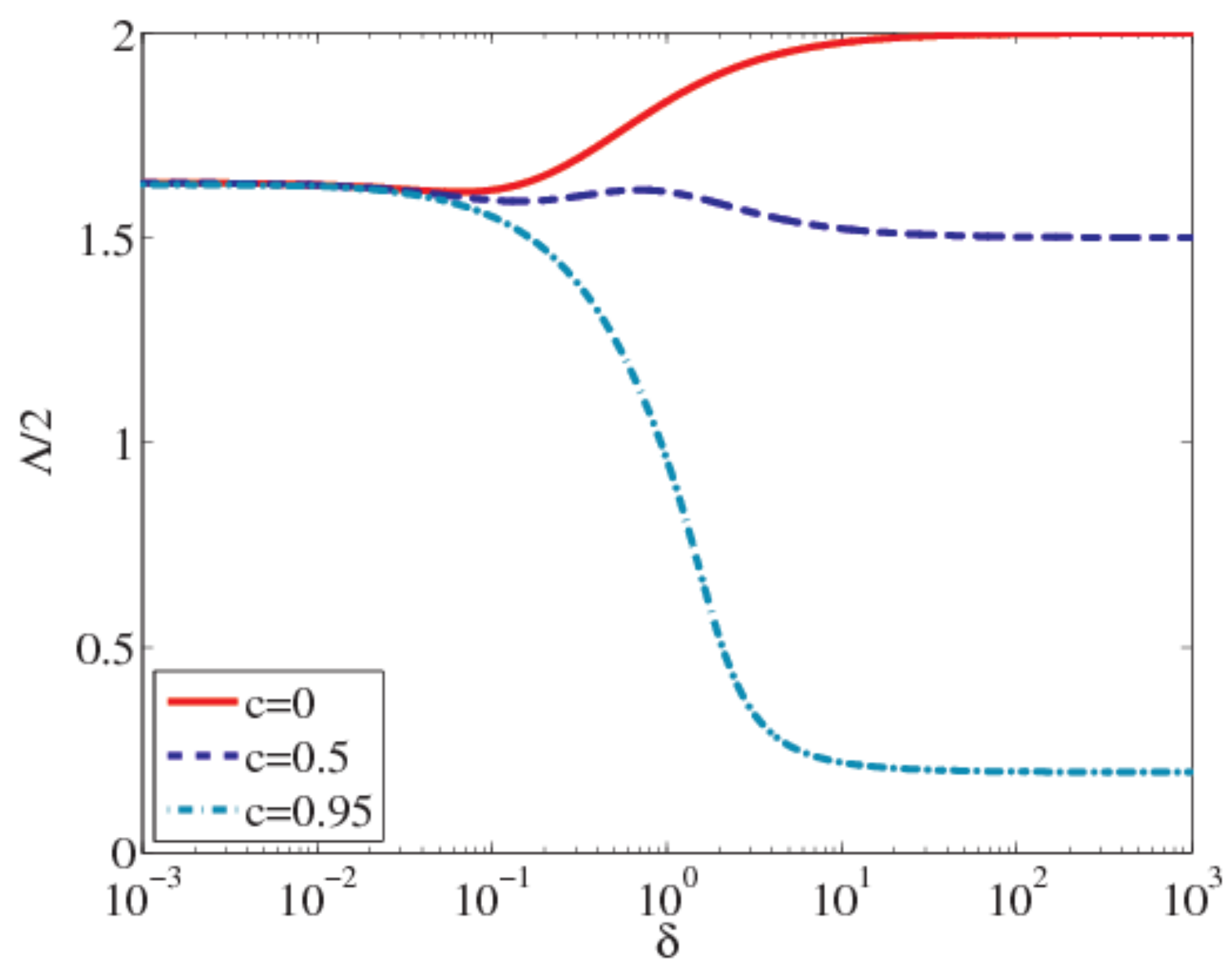}}
\subfigure[$\,w=5$]{\includegraphics[width=0.32\textwidth]{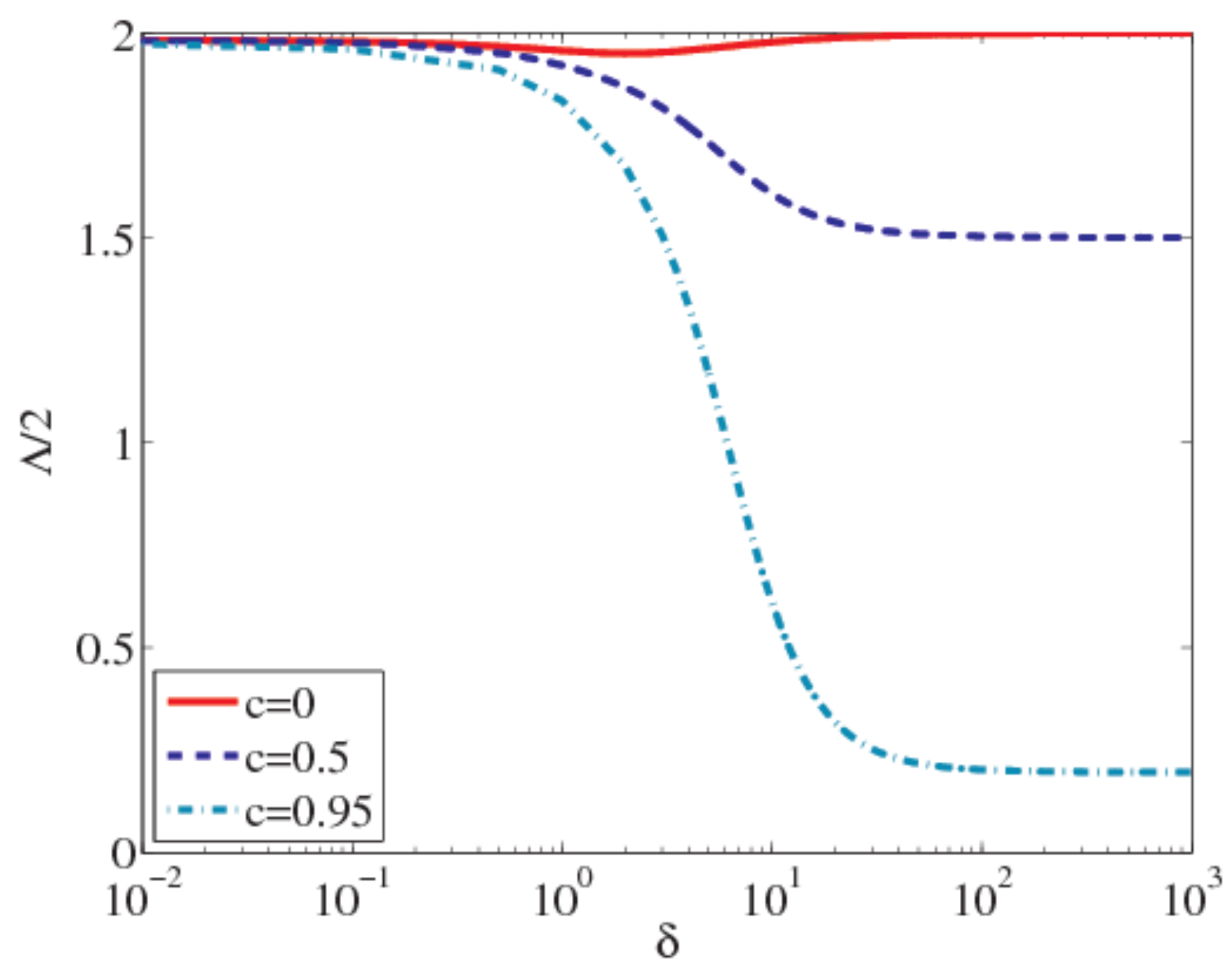}}
\end{center}
\caption{Parameter study of the mean growth rate $\langle\Lambda\rangle$: it is positive for all positive values of $\delta$.  For $w=0$, $\langle\Lambda\rangle$ approaches zero as $\delta\rightarrow 0$.  For non-zero $w$-values, $\langle\Lambda\rangle$ approaches positive limiting values as $\delta\rightarrow 0$ and as $\delta\rightarrow\infty$.}
\label{fig:param}
\end{figure}

To study the general case, we turn to the numerical results in Fig.~\ref{fig:param}, where we focus on three parameters, $w/D_Y$, $\delta$, and $\kcorr$ (we set $D_Y=1$).  
The mean growth rate $\langle\Lambda\rangle$ is positive in all of the parameter studies.  For $w=0$, it tends to zero as $\delta\rightarrow 0$.  For other $w$-values, it approaches definite, positive limiting values as $\delta\rightarrow 0$ and $\delta\rightarrow\infty$.  The zero-limit is $\kcorr$-independent, while the $\delta\rightarrow\infty$ limit depends on $\kcorr$: the stronger the correlation, the smaller the asymptotic value taken by $\langle\Lambda\rangle$.  We now compare the analytical results of the model with numerically-simulated flows.

\section{The random-phase sine flow}
\label{sec:sineflow}
In this section we compute the orientation statistics for the random-phase sine flow~\cite{Pierrehumbert1994,Antonsen1996}.  This flow is amenable to a comparison with our stochastic model because the randomisation of the phases breaks the invariant tori, induces homogeneity, and promotes mixing~\cite{ONaraigh2007}.  This is in contrast to the unrandomised sine flow studied elsewhere~\cite{Gonzalez2010}.

The flow is quasi-periodic and is defined as follows.  In the $j^{\mathrm{th}}$ period,
\begin{subequations}
\begin{equation}
u=\ampl\sin\left(ky+\phi_j\right),\qquad v=0,
\end{equation}
for the first half-period, and
\begin{equation}
u=0,\qquad v=\ampl\sin\left(kx+\psi_j\right)
\end{equation}%
\label{eq:sineflow}%
\end{subequations}%
for the second. Here $\tau$ is the period of the flow, $A_0$ is the flow amplitude, and $\phi_j$ and $\psi_j$ are random phases. 

In the $j^\mathrm{th}$ cycle of the sine flow, the strain and the vorticity have the form
\begin{subequations}
%\begin{multline}
\begin{equation}
d=\tfrac{1}{2}\ampl kH_\tau\left(t\right)\cos\left(ky+\phi_j\right)+\\
\tfrac{1}{2}\ampl \left(1-H_\tau\left(t\right)\right)\cos\left(kx+\psi_j\right),
\end{equation}
%\end{multline}
%\begin{multline}
\begin{equation}
\omega=A_0kH_\tau\left(t\right)\cos\left(ky+\phi_j\right)+\\
A_0k\left(1-H_\tau\left(t\right)\right)\cos\left(kx+\psi_j\right),
\end{equation}
%\end{multline}
\label{eq:om_d}%
\end{subequations}%
where $H_\tau\left(t\right)$ is equal to unity in the first half-period and is zero in the second.
Equations~\eqref{eq:om_d} imply that $d$ and $\omega$ are not correlated with each other, and that their autocorrelation functions are nonzero only in a half-period, wherein they take constant values.  Thus, to mimic the stirring protocol~\eqref{eq:sineflow} by an OU process, we identify the period $\tau$ of the sine flow with the OU decay timescale, and take $w=\kcorr=0$ and $D_Y=D_Z=A_0^2k^2/\left(2\tau\right)$ in Eq.~\eqref{eq:model_ou}.
Furthermore, since $\angleS=\mathrm{Const.}$ for the sine flow,
\begin{equation}
\left(\partial_t+\bm{u}\cdot\nabla\right)\xs
%=2\left(\partial_t+\bm{u}\cdot\nabla\right)\beta
=\omega-2d\cos\xs,
\label{eq:gamma_sineflow}%
\end{equation}%
where $\xs=2\left(\beta-\angleS+\tfrac{1}{4}\pi\right)=2\beta$.
Using Eqs.~\eqref{eq:om_d}, this becomes
\begin{equation}
\frac{d\xs}{1+\cos\xs}=-\ampl k\cos\left(ky_n+\phi_n\right)
\end{equation}
in the first half-period, and
\begin{equation}
\frac{d\xs}{1-\cos\xs}=\ampl k\cos\left(kx_{n+1}+\psi_n\right)
\end{equation}
in the second.
The growth rate of the tracer gradient along a trajectory originating at $\bm{x}_0$ is thus
%
%
%\begin{multline}
\begin{equation}
\Lambda\left(\bm{x}_0\right)=\\
\lim_{N\rightarrow\infty}\frac{1}{N}\sum_{n=0}^N\left[-d^{n+1/2}\sin\xs^{n+1/2}-d^{n+1}\sin\xs^{n+1}\right].
\label{eq:lambda_traj}
\end{equation}
%\end{multline}
%
%

\begin{figure}
\begin{center}
\subfigure[$\,A_0=0.5$]{\includegraphics[width=0.32\textwidth,height=0.25\textwidth]{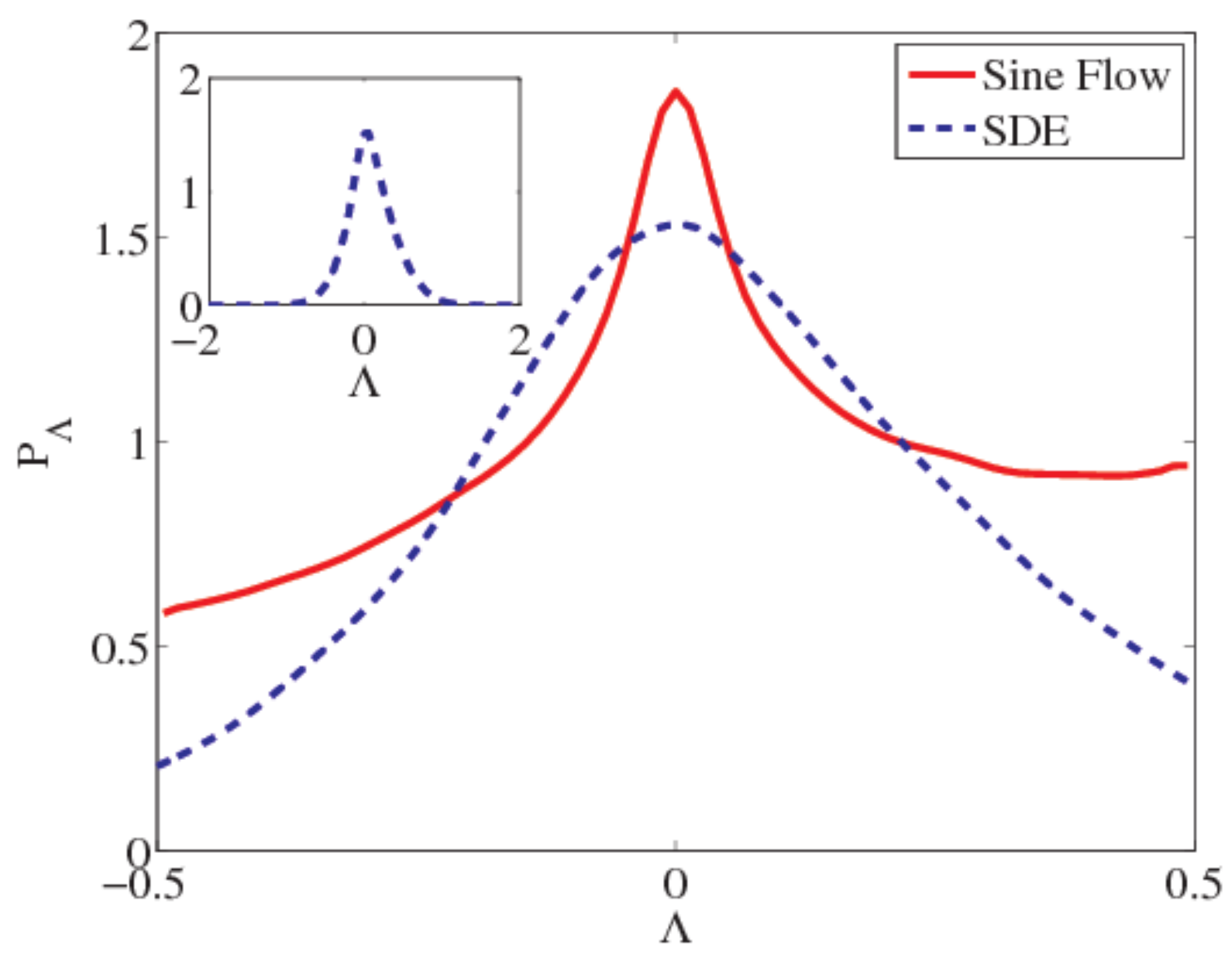}}
\subfigure[$\,A_0=1$]{\includegraphics[width=0.32\textwidth,height=0.25\textwidth]{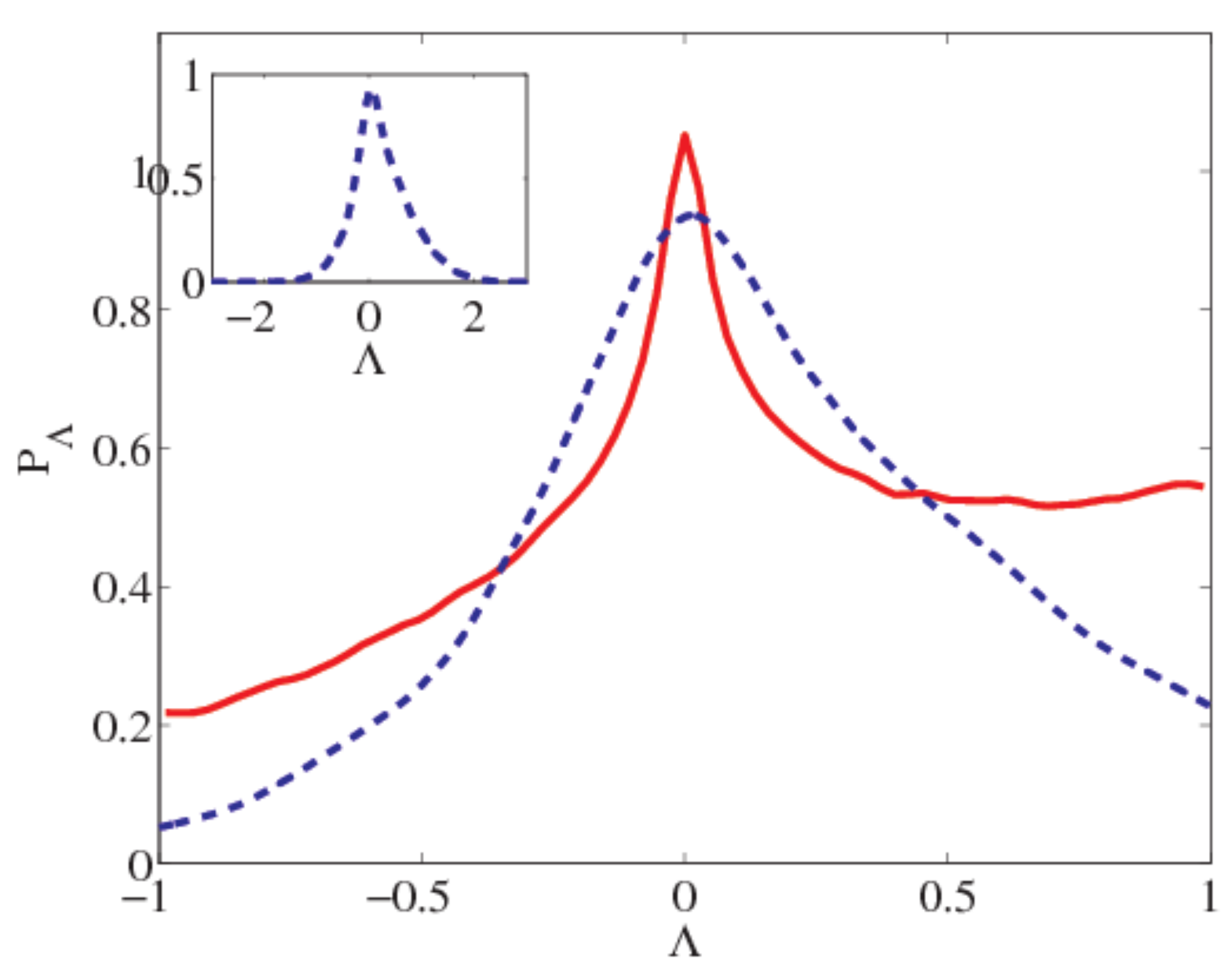}}
\subfigure[$\,A_0=2$]{\includegraphics[width=0.32\textwidth,height=0.25\textwidth]{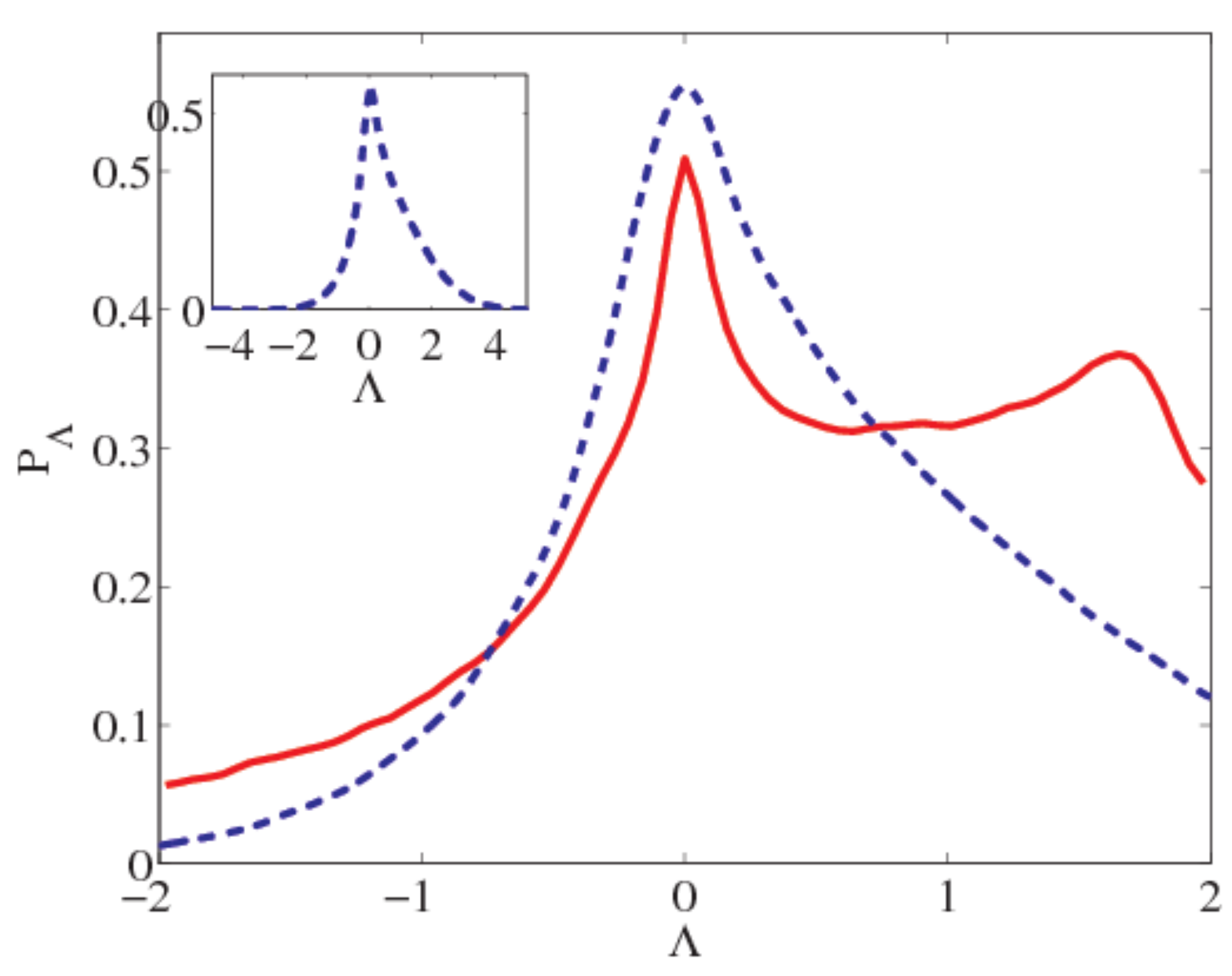}}
\end{center}
\caption{PDF of the growth rate $\Lambda$ as a function of flow amplitude $A_0$.  In the main figures, we show a comparison between the OU model and the sine flow, while the insets contain the PDF of the OU model over an extended range.  In the sine-flow case, the PDF takes values in a bounded interval, while the tails of the OU model extend to $\pm\infty$.}
\label{fig:pdf_sineflow1}
\end{figure}

\begin{figure}
\begin{center}
\subfigure[$\,A_0=0.5$]{\includegraphics[width=0.32\textwidth]{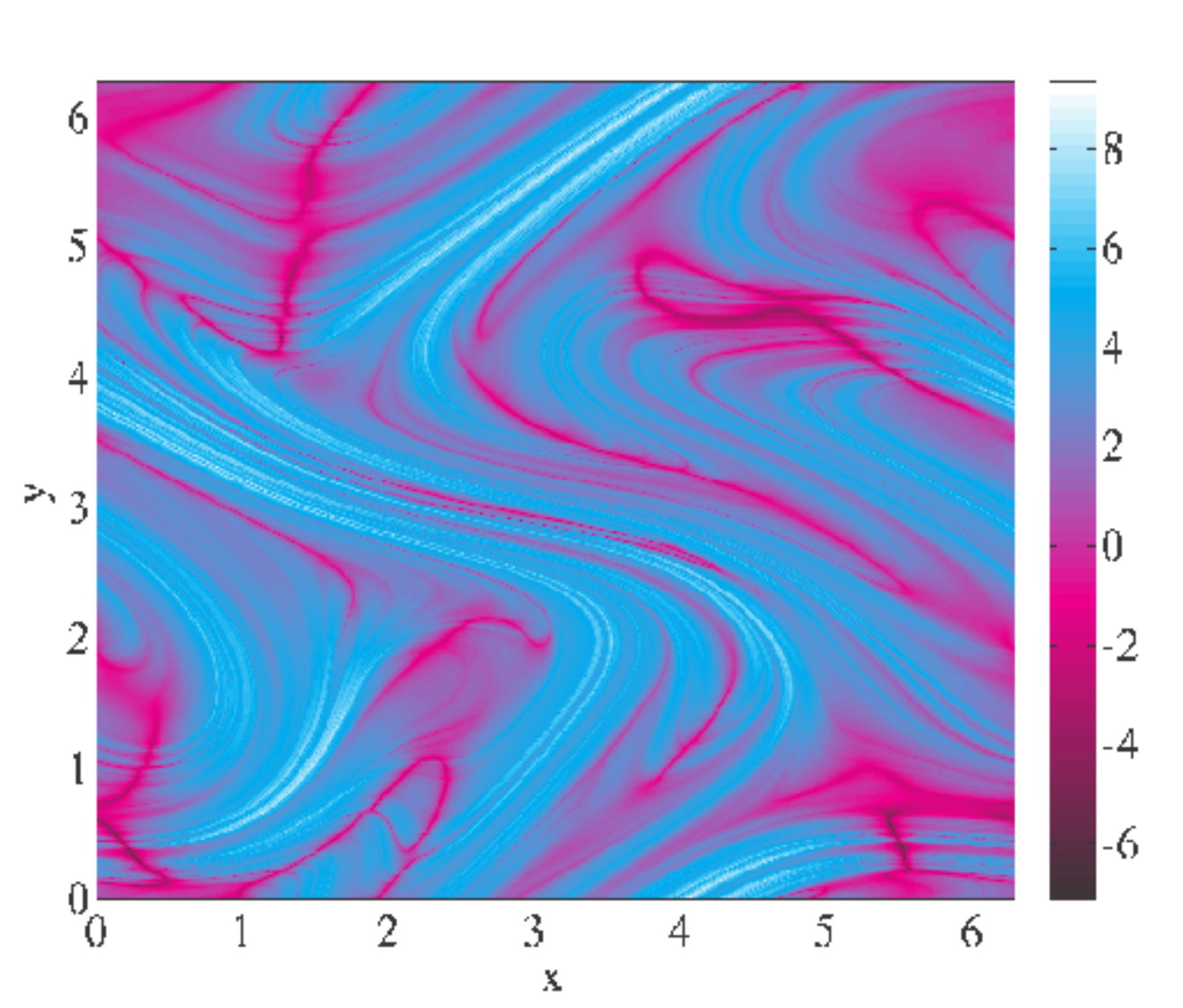}}
\subfigure[$\,A_0=1$]{\includegraphics[width=0.32\textwidth]{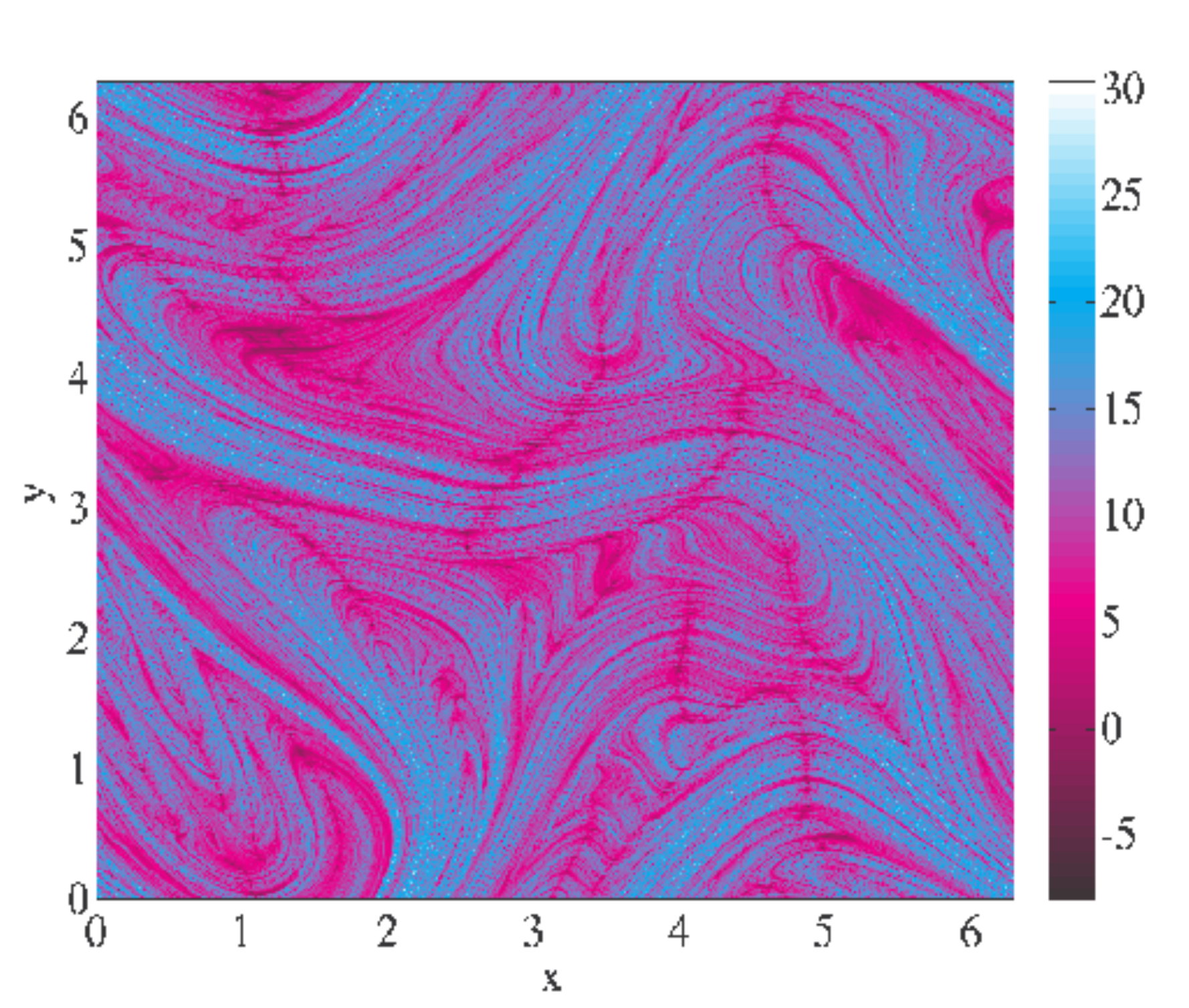}}
\subfigure[$\,A_0=2$]{\includegraphics[width=0.32\textwidth]{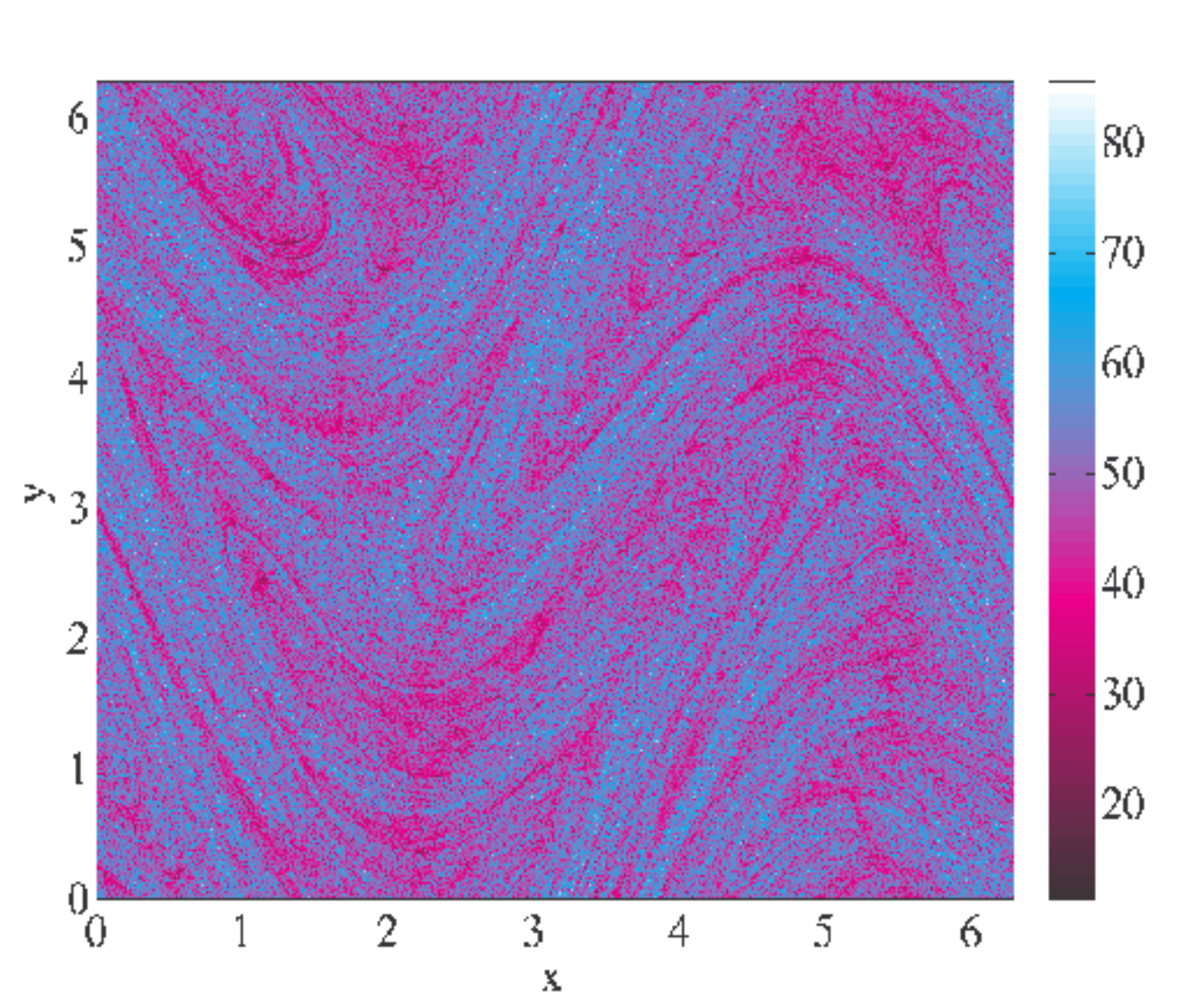}}
\end{center}
\caption{The growth rate of the tracer gradient along trajectories, according to Eq.~\eqref{eq:lambda_traj}.}
\label{fig:ftle_sineflow}
\end{figure}

We examine the PDF of angles $X$ as generated by the sine flow 
\begin{figure}
\begin{center}
\subfigure[$\,A_0=0.5$]{\includegraphics[width=0.32\textwidth]{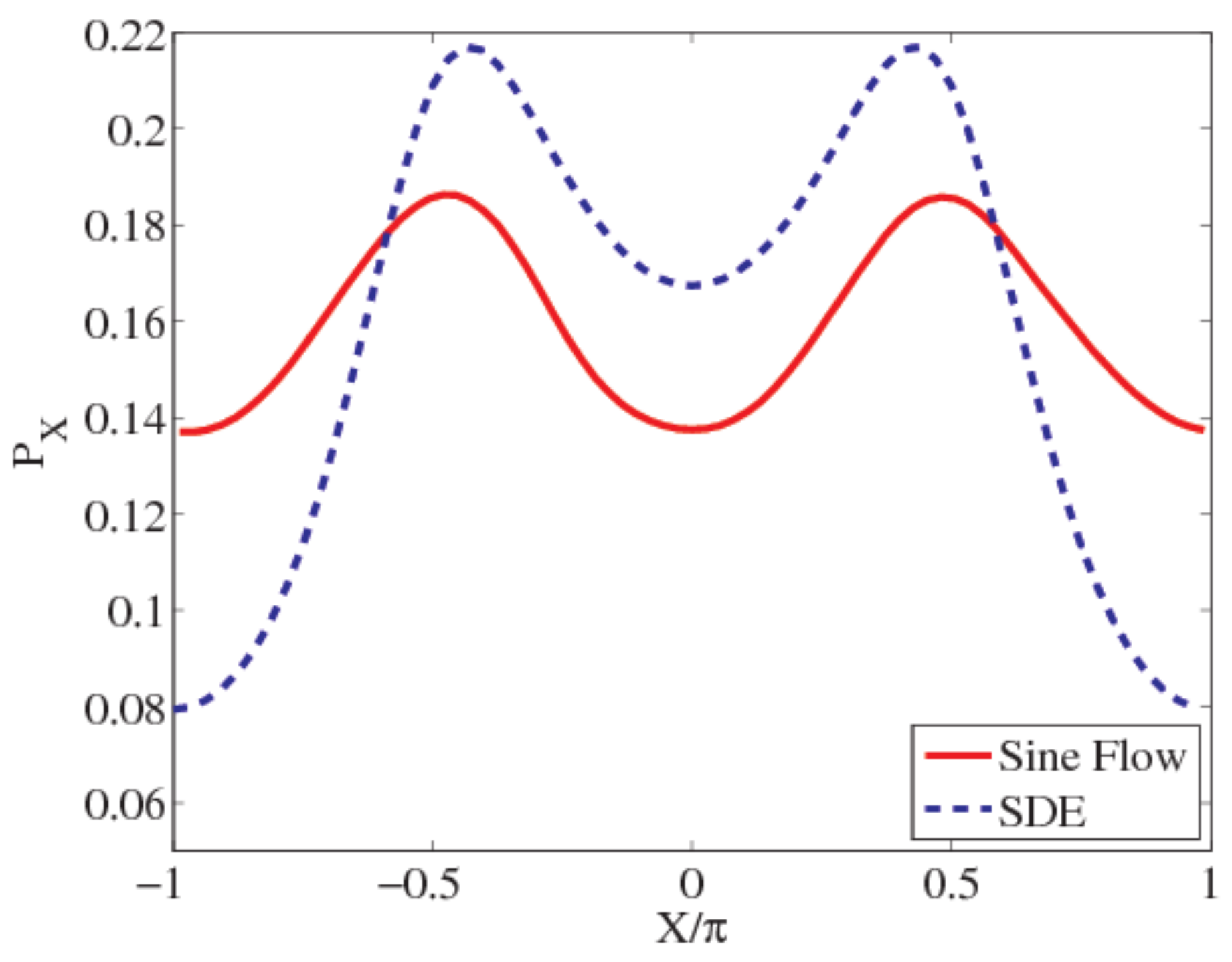}}
\subfigure[$\,A_0=1$]{\includegraphics[width=0.32\textwidth]{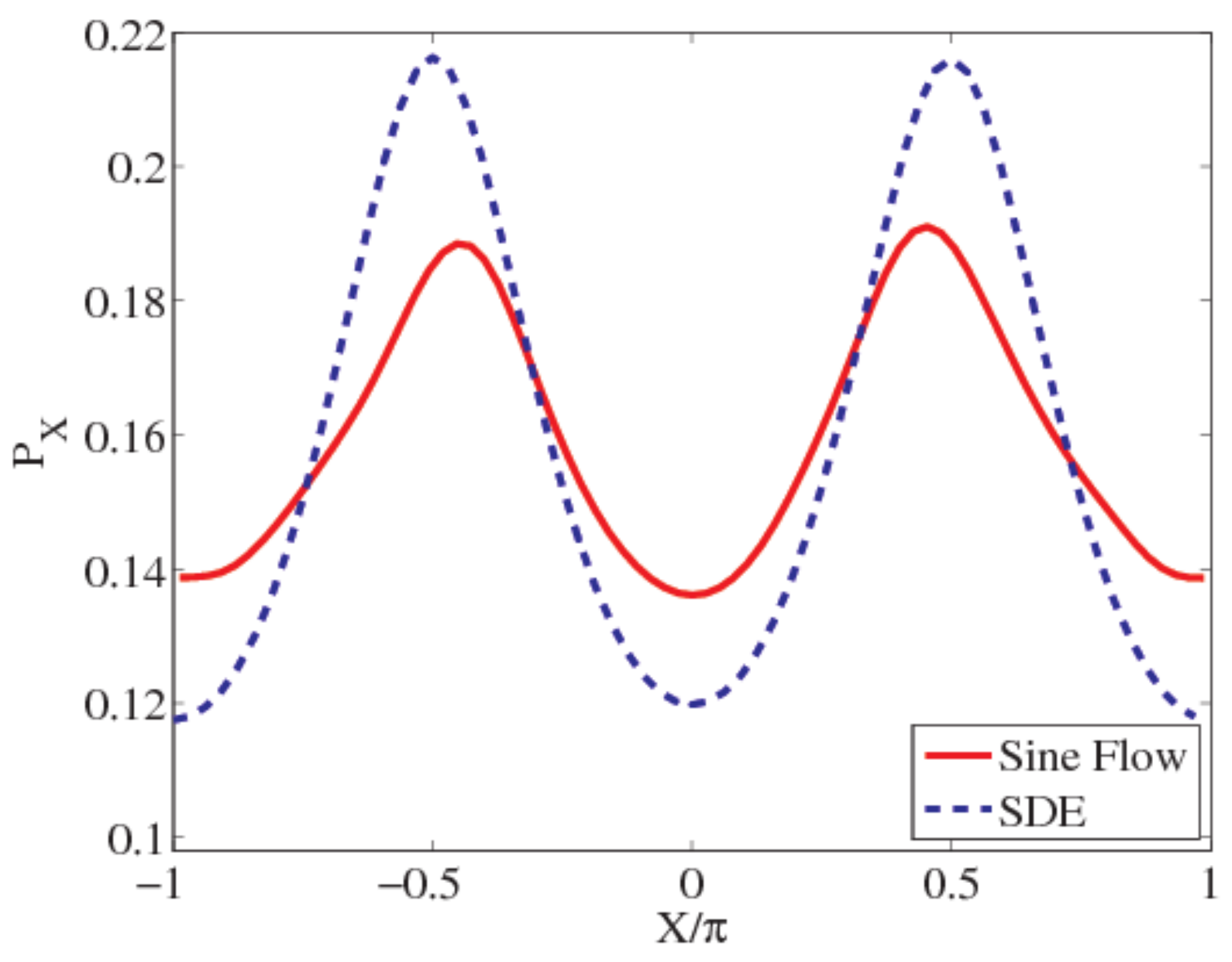}}
\subfigure[$\,A_0=2$]{\includegraphics[width=0.32\textwidth]{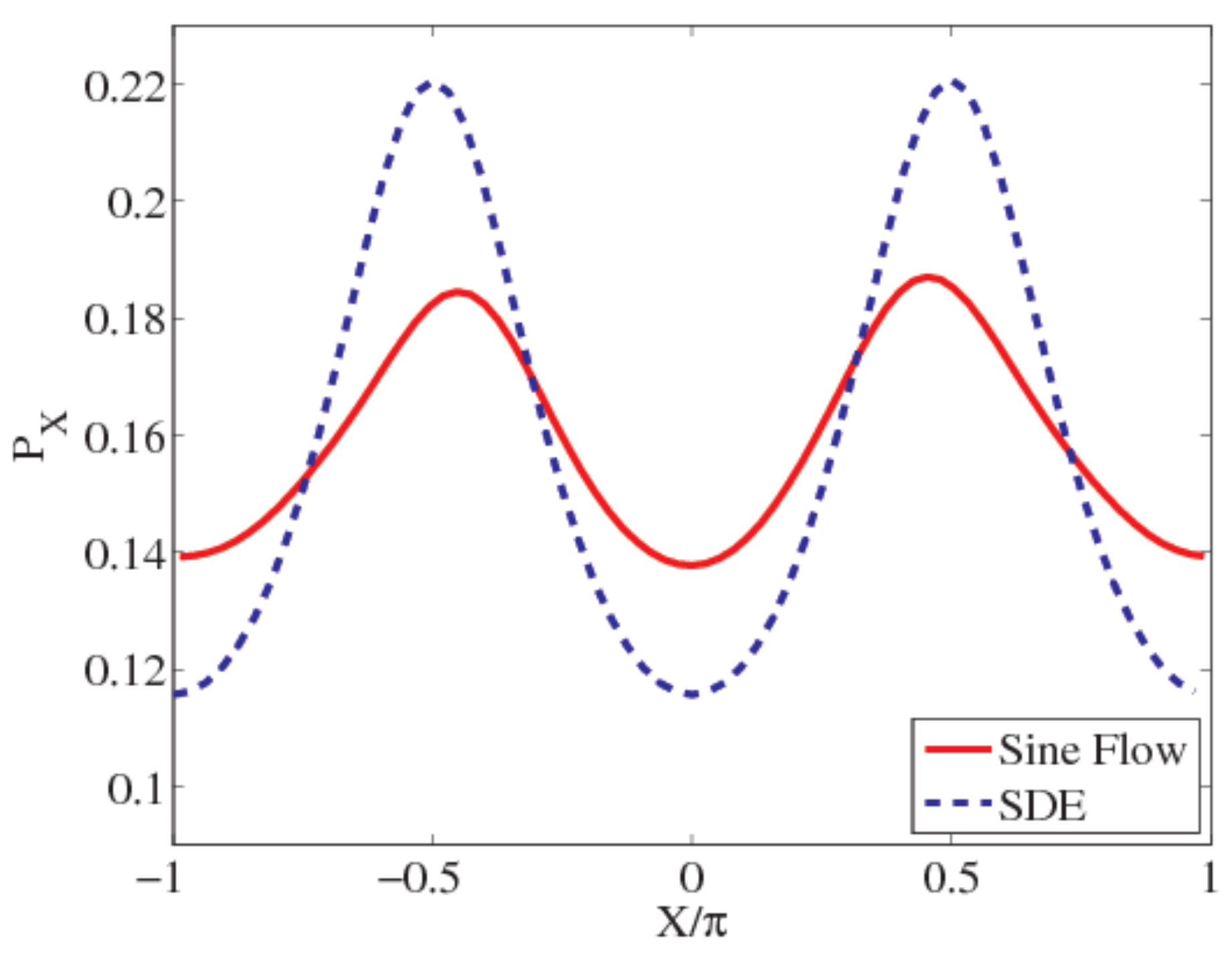}}
\end{center}
\caption{PDF of the angle $X$ as a function of flow amplitude $A_0$: comparison between the sine flow and the OU model.  The comparison yields good qualitative agreement, despite the difference between the statistics of the underlying forcing terms in each model. }
\label{fig:pdf_sineflow}
\end{figure}
and compare it with the PDF obtained from the model OU process (Fig.~\ref{fig:pdf_sineflow}).  The results are qualitatively similar: in both cases, the PDF is symmetric about $X=0$, with maxima close to $X=\pm\pi/2$.  The sine-flow maxima deviate slightly from these values: this is probably due to the slow convergence of the PDF as the number of particles in the ensemble is increased.  In the OU case, the FP equation gives rise to a fully-converged PDF, whose maxima occur at exactly $X=\pm \pi/2$. 
% 
% while in the sine-flow case, the maxima occur at $|X|>\pi/2$ for small $A_0$, and at $|X|<\pi/2$ for larger values of % $A_0$.  
%
%
This qualitative similarity between the two PDFs is good, in view of the radically different noise-generating processes in each model: in the OU case, the noise terms are unbounded, while in the sine-flow case the fluctuations in $Y$ and $Z$ are bounded,   $|Y|, |Z|\leq A_0k$.  We also compute the PDF associated with the growth rate (Fig.~\ref{fig:pdf_sineflow1}).
This pair of PDFs for the sine flow demonstrates that on average, the tracer gradient aligns with the compressive direction of the flow.
Finally, we examine the Eulerian structure of the flow by plotting the growth rate $\Lambda\left(\bm{x}_0\right)$ (Fig.~\ref{fig:ftle_sineflow}).  This is 
spatially homogeneous, thus underscoring the efficient nature of the random-phase sine flow in mixing the passive tracer.  It also provides further justification for our application of the OU model to the problem in hand, since the flow statistics are, on average, the same along all trajectories.

\section{Forced two-dimensional turbulence}
\label{sec:turb}

In this section we compare the OU model of the orientation dynamics with the results of numerical simulations of forced two-dimensional turbulence.  We present results for the solution of the vorticity equation
\begin{equation}
\frac{\partial\omega}{\partial t}+\bm{u}\cdot\nabla\omega=-\left(-1\right)^p\nu_p\nabla^p\omega+Q-\nu_0\omega,
\label{eq:vorticity}
\end{equation}
where $Q$ is a forcing term and $-\nu_0\omega$ is a damping term that prevents a buildup of energy at large scales.
(Simulations for Rayleigh damping~\cite{DelSole1995} have also been carried out, and yield similar results.)
To obtain a stochastic driving force on a particular scale $k_\mathrm{e}$, we use the method developed by Lilly~\cite{Lilly1969} (see also~\cite{Molenaar2004}).  The combination of damping and forcing in Eq.~\eqref{eq:vorticity} yields a statistically-steady state.  Moreover,  the tracer, driven by the flow~\eqref{eq:vorticity}, also reaches a steady state, in the sense that $\theta/\|\theta\|_2$ exhibits the so-called strange eigenmode~\cite{Pierrehumbert1994,Haller2003}.  Thus, it is legitimate to regard the late-time statistics of the angle $\beta$ as being drawn from a stationary distribution.

\begin{figure}
\begin{center}
\subfigure[]{\includegraphics[width=0.32\textwidth]{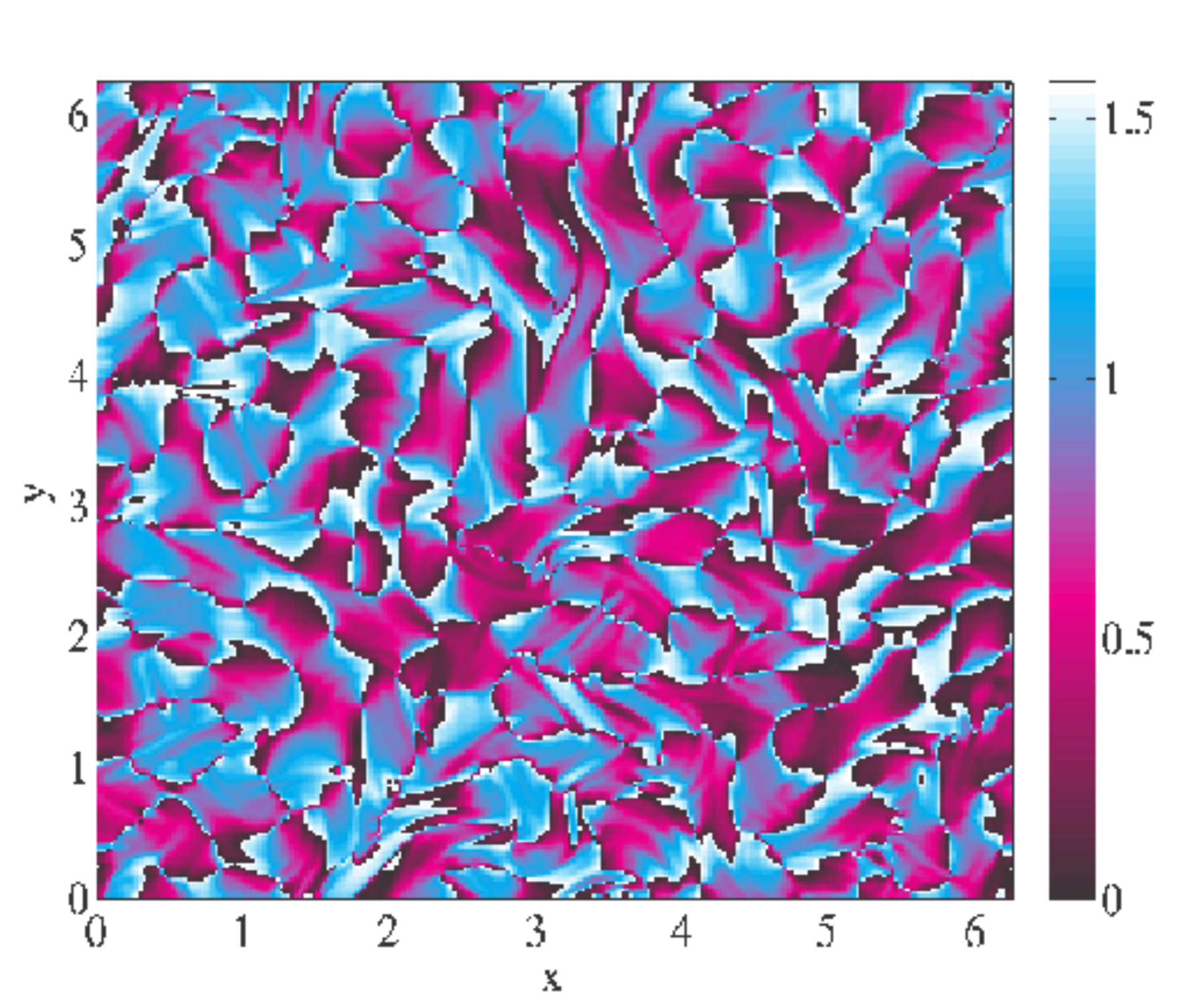}}
\subfigure[]{\includegraphics[width=0.32\textwidth]{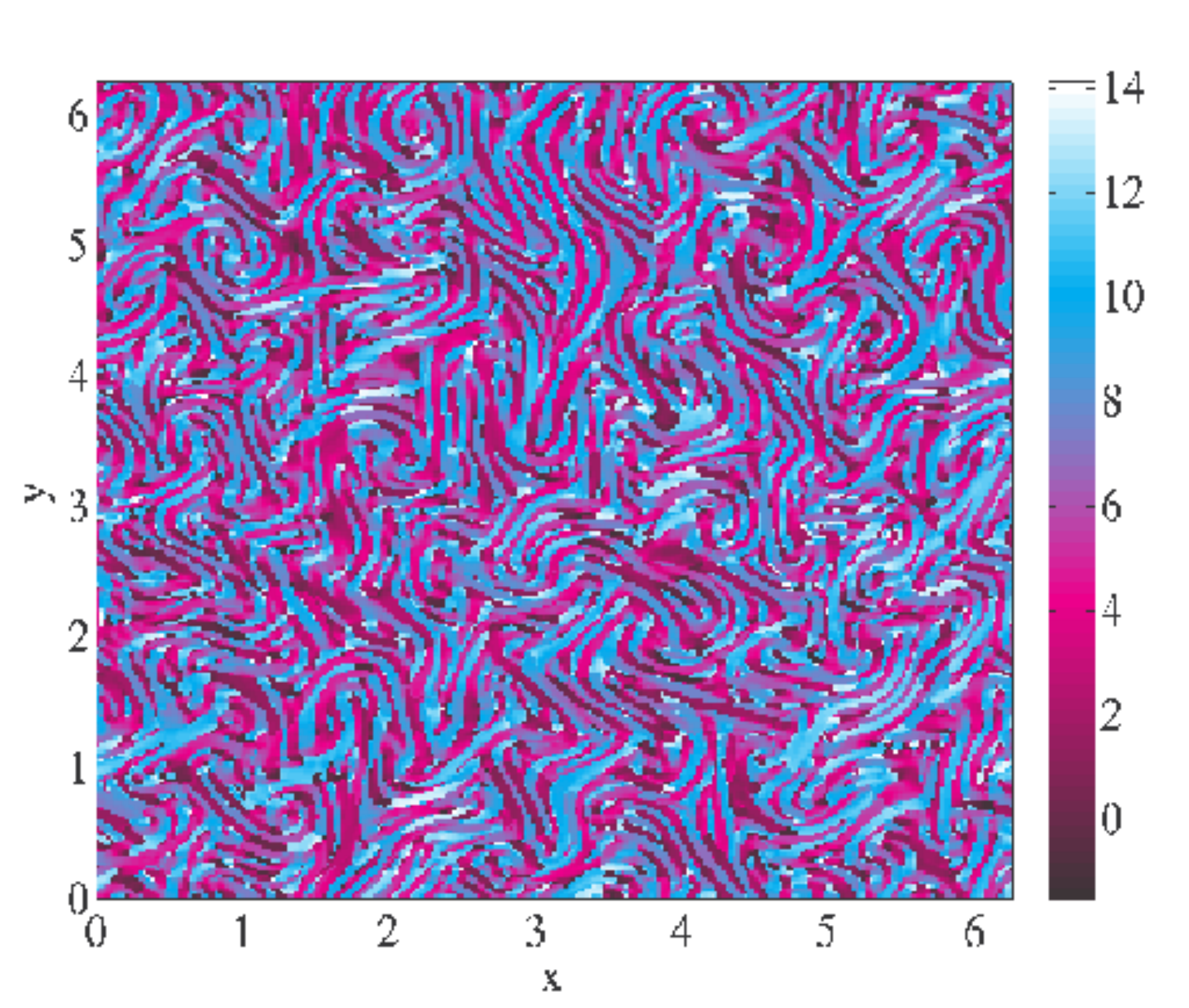}}
\subfigure[]{\includegraphics[width=0.32\textwidth]{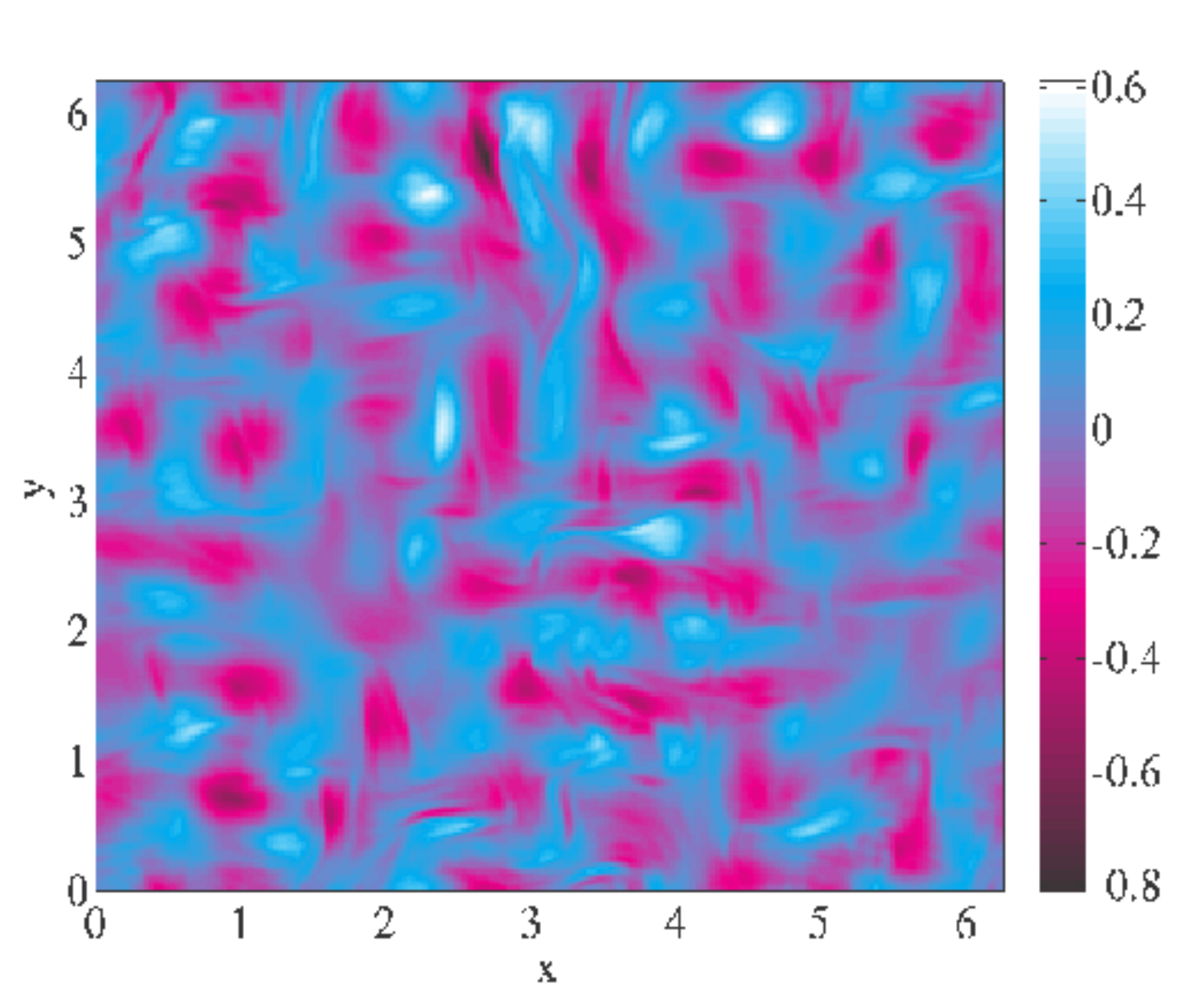}}
\end{center}
\caption{Orientation dynamics of forced two-dimensional turbulence in a statistically steady state: shapshots of (a) the angle $\angleS$; (b) the angle $X$; (c) the strain rate $d$.}
\label{fig:vorticity_pictures1}
\end{figure}

\begin{figure}
\begin{center}
\subfigure[]{\includegraphics[width=0.454\textwidth]{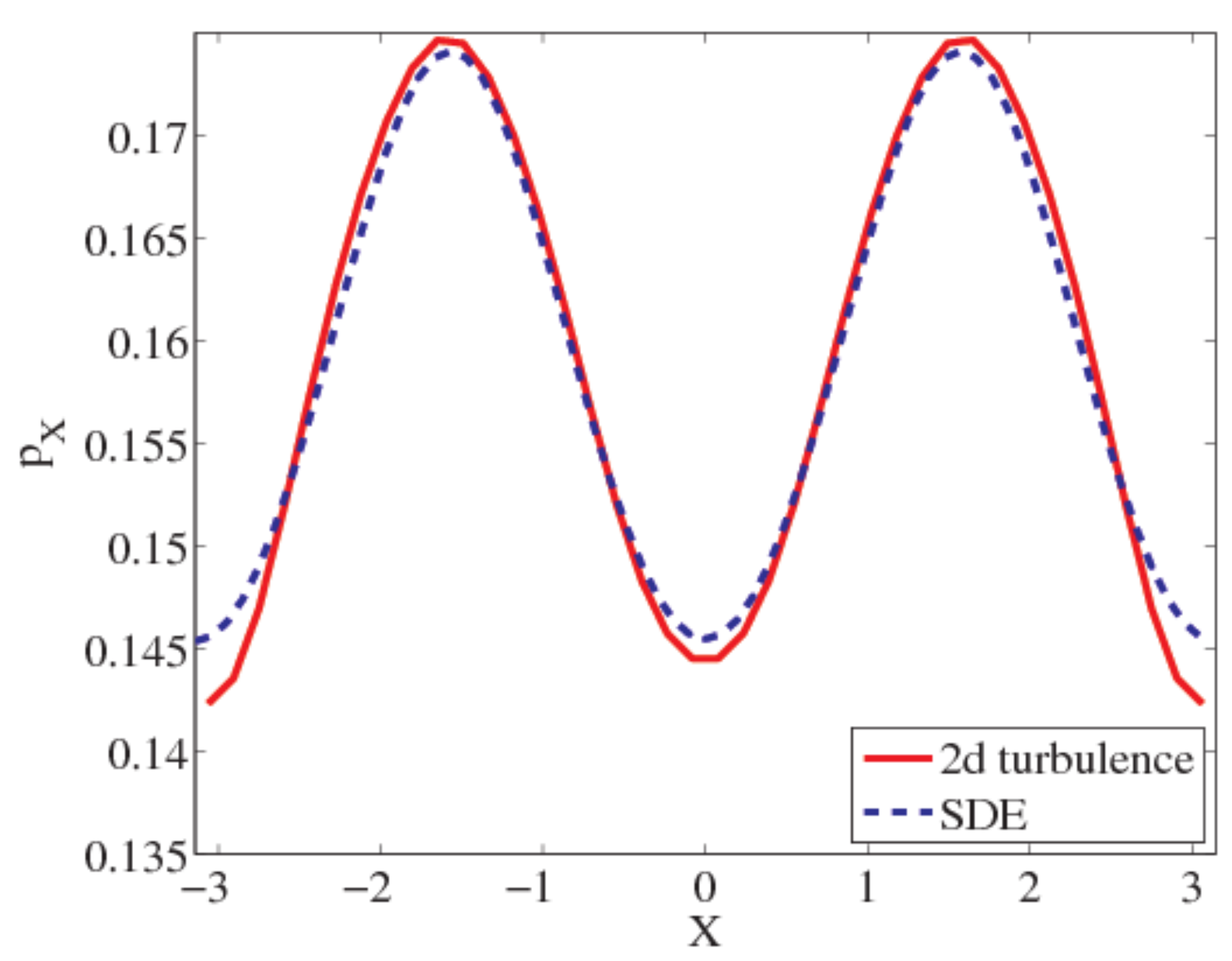}}
\subfigure[]{\includegraphics[width=0.45\textwidth]{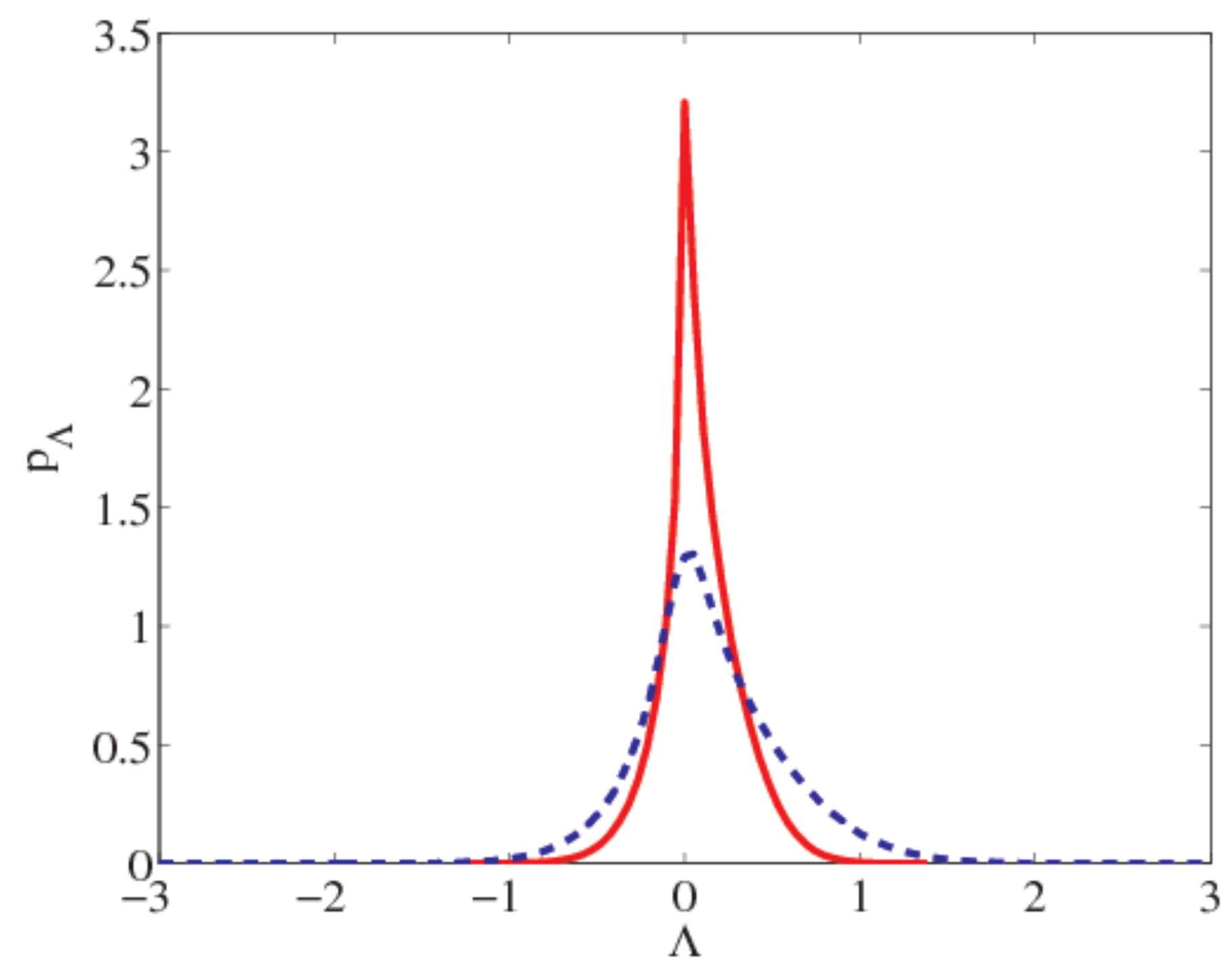}}
\end{center}
\caption{(a) The PDF of the angle $X$ according to the 2d turbulence simulation (solid line), and the OU model (broken line) (b) The PDF of the growth rate.  The model is in excellent qualitative agreement with the turbulence.}
\label{fig:turb_results}
\end{figure}

To map the flow problem onto the OU dynamics, we examine the Eulerian fields
\[
\tilde{w}=\tfrac{1}{2}\omega+\partial_t\angleS+\bm{u}\cdot\nabla\angleS,\qquad
\tilde{\lambda}=\mathrm{sign}\left(d\right)\sqrt{d^2+s^2}.
\]
A snapshot of these fields at the same time is shown in Fig.~\ref{fig:vorticity_pictures1}.  We solve Eq.~\eqref{eq:vorticity} using a standard, semi-implicit pseudospectral method.  We take the order of the viscosity to be $p=8$; we also take $\nu_p=5.9\times 10^{-30}$ in both the $\omega$- and the $\theta$-equations, and $\nu_0=0.05$.  The spatial resolution is $N^2=256^2$ and the timestep is set to $\Delta t=10^{-3}$.
We identify the decay time $\tau$ with $\nu_0^{-1}$ and examine Eulerian time series of $\tilde{w}$ and $\tilde{\lambda}$ over the steady-state period of the numerical integration.  These scalar variables fluctuate around certain mean values, such that $\langle \|\tilde{\lambda}\|_2^2\rangle_t=0.05/\tau$,  $\langle\|\tilde{w}\|_2^2\rangle_t=0.94/\tau$, and 
$\langle \int \mathd^2x\tilde{\lambda}\tilde{w}\rangle_t=0.00$ (the brackets $\langle\cdot\rangle_t$ denote temporal averaging over a long steady-state interval).  Thus, we model the orientation dynamics as OU processes with parameters $D_Y=0.05/\tau$, $D_Z=0.94/\tau$, $\kcorr=0$, and $w=0$. Making these identifications, we compare the PDF of the angle $X$ and the growth rate $\Lambda$, as generated both by the model OU process, and the two-dimensional turbulence.
The results are shown in Fig.~\ref{fig:turb_results}.
The angle PDFs for the turbulence and for the OU model are in very close agreement, with large maxima at exactly $X=\pm\pi/2$.  The PDFs for the growth rate are in close qualitative agreement: the curve is asymmetric around $\Lambda=0$, and is skewed towards positive values.  Both curves fall off sharply away from $\Lambda=0$.  The first moment of the turbulence PDF is less than that of the OU process ($\Lambda_{1,OU}=0.15$; $\Lambda_{1,TURB}=0.08$), and the tail of the turbulence PDF is less `fat' than its OU counterpart.
Nevertheless, the close qualitative agreement between the model and the flow both in this section and in Sec.~\ref{sec:sineflow} confirms the usefulness of the OU description in modelling the growth rate of a tracer gradient under rapid stirring.

\section{Conclusions}
\label{sec:conc}

We have formulated a stochastic model of the orientation dynamics of a tracer gradient in a manner designed to mimic 
mixing under externally-imposed rapidly-varying stirring in two dimensions.
The model consists of a set of SDEs for the angle $\zeta$, and for the forces that act thereon.  
The model is a general one that includes the adiabatic description of Lapeyre~\cite{Lapeyre1999} as a special case.  However, we focus entirely on the rapid case: we analyse the SDEs using the Fokker--Planck (FP) equation and compute the distrubtion of orientation angles and growth rates.  The use of the FP equation gives rise to some analytical results in the limit of zero correlation time, and in the special case when the mean vorticity and the correlation between the forcing terms both vanish: the tracer gradient aligns with the compressional direction, and the mean growth rate of the tracer gradient is positive and given by an explicit expression in $\delta=D_Z/D_Y$ involving elliptic functions.  For finite correlation times, numerical solutions suggest that the PDF is always skewed to positive values, giving rise to a positive mean growth rate.
%
%
%Because we are interested in rapidly-varying flows, 
We compare our model with two rapid-flow protocols: the random-phase sine flow, and forced two-dimensional turbulence.  The qualitative agreement between the model PDFs and the PDFs generated from the flow protocols is excellent, and confirms the validity of the approach.  Since our stochastic model is amenable both to mathematical and numerical analysis (the latter in minutes on a desktop computer), we anticipate its generalisation to three-dimensional flows.

\subsection*{Acknowledgements}

The author wishes to thank M. Bustamante, D. D. Holm and J.-L. Thiffeault for helpful conversations.

%\bibliographystyle{unsrt}
%\bibliography{2dturbulence_bibliography}

\begin{thebibliography}{10}

\bibitem{Falkovich2001}
G.~Falkovich, K.~Gawedzki, and M.~Vergassola.
\newblock Particles and fields in fluid turbulence.
\newblock {\em Rev. Mod. Phys.}, 73:913--975, 2001.

\bibitem{Kraichnan1974}
R.~Kraichnan.
\newblock Convection of a passive scalar by a quasi-uniform random straining
  field.
\newblock {\em J. Fluid Mech.}, 64:737--762, 1974.

\bibitem{Pierrehumbert1994}
R.~T. Pierrehumbert.
\newblock Tracer microstructure in the large-eddy dominated regime.
\newblock {\em Chaos, Solitons and Fractals}, pages 1091--1110, 1994.

\bibitem{Schekochihin2004}
A.~A. Schekochihin, P.~H. Haynes, and S.~C. Cowley.
\newblock Diffusion of passive scalar in a finite-scale random flow.
\newblock {\em Phys. Rev. E}, 70:046304, 2004.

\bibitem{Balkovsky1999}
E.~Balkovsky and A.~Fouxon.
\newblock {Universal long-time properties of Lagrangian statistics in the
  Batchelor regime and their application to the passive-scalar problem}.
\newblock {\em Phys. Rev. E}, 60:4164--4174, 1999.

\bibitem{Bernard1998}
D.~Bernard, K.~Gawedzki, and Kupiainen.
\newblock Slow modes in passive advection.
\newblock {\em J. Stat. Phys}, 909:519, 1998.

\bibitem{Lapeyre1999}
G.~Lapeyre, P.~Klein, and B.~L. Hua.
\newblock Does the tracer gradient vector align with the strain eigenvectors in
  2d turbulence?
\newblock {\em Phys. Fluids}, 11:3729--3737, 1999.

\bibitem{Weiss1991}
J.~Weiss.
\newblock The dynamics of enstrophy transfer in two-dimensional hydrodynamics.
\newblock {\em Physica D}, 48:273, 1991.

\bibitem{Lapeyre2002}
G.~Lapeyre.
\newblock {Characterization of finite-time Lyapunov exponents and vectors in
  two-dimensional turbulence}.
\newblock {\em Chaos}, 12:688--698, 2002.

\bibitem{Gonzalez2010}
M.~Gonzalez and P.~Parantho\"en.
\newblock On the role of unsteady forcing of tracer gradient in local stirring.
\newblock {\em Eur. J. Mech. B}, 29:143--152, 2010.

\bibitem{Garcia2005}
A.~Garcia, M.~Gonzalez, and P.~Parantho\"en.
\newblock On the alignment dynamics of a passive scalar gradient in a
  two-dimensional flow.
\newblock {\em Phys. Fluids}, 17:117102, 2005.

\bibitem{Gonzalez2009}
M.~Gonzalez.
\newblock {Kinematic properties of passive scalar gradient predicted by a
  stochastic Lagrangian model}.
\newblock {\em Phys. Fluids}, 21:055104, 2009.

\bibitem{RiskenBook}
H.~Risken.
\newblock {\em {The Fokker--Planck Equation}}.
\newblock Springer, second edition, 1989.

\bibitem{Dresselhaus1992}
E.~Dresselhaus and M.~Tabor.
\newblock The kinematics of stretching and alignment of material elements in
  general flow fields.
\newblock {\em J. Fluid Mech.}, 236:415--444, 1992.

\bibitem{Holm2010}
J.~D. Gibbon and D.~D. Holm.
\newblock The dynamics of the gradient of potential vorticity.
\newblock {\em J. Phys. A: Math. Theor.}, 43:172001, 2010.

\bibitem{Molenaar2004}
D.~Molenaar, H.~J.~H. Clercx, and G.~J.~F. van Heijst.
\newblock Angular momentum of forced 2d turbulence in a square no-slip domain.
\newblock {\em Physica D}, page 329–340, 2004.

\bibitem{Li1998}
J.~H. Li and Z.~Q. Huang.
\newblock Transport of particles caused by correlation between additive and
  multiplicative noise.
\newblock {\em Phys. Rev. E}, 57:3917--3922, 1998.

\bibitem{BaroneBook}
A.~Barone and G.~Patern\`{o}.
\newblock {\em Physics and applications of the Josephson effect}.
\newblock Wiley, 1982.

\bibitem{KonotopBook}
V.~V. Konotop and L.~V\'azquez.
\newblock {\em Nonlinear random waves}.
\newblock World Scientific Publishing, 1994.

\bibitem{Antonsen1996}
Jr. T.~M.~Antonsen, Z.~Fan, E.~Ott, and E.~Garcia-Lopez.
\newblock The role of chaotic orbits in the determination of power spectra.
\newblock {\em Phys. Fluids}, 8:3094--3104, 1996.

\bibitem{ONaraigh2007}
L.~\'O N\'araigh and J.-L. Thiffeault.
\newblock Bubbles and filaments: Stirring a cahn-hilliard fluid.
\newblock {\em Phys. Rev. E}, 75:016216, 2007.

\bibitem{DelSole1995}
T.~DelSole and B.~F. Farrell.
\newblock A stochastically excited linear system as a model for
  quasigeostrophic turbulence: Analytic results for one- and two-layer fluids.
\newblock {\em Journal of the Atmospheric Sciences}, 52:2531--2547, 1995.

\bibitem{Lilly1969}
D.~K. Lilly.
\newblock Numerical simulation of two-dimensional turbulence.
\newblock {\em Phys. Fluids II}, page 240–249, 1969.

\bibitem{Haller2003}
W.~Liu and G.~Haller.
\newblock Strange eigenmodes and decay of variance in the mixing of diffusive
  tracers.
\newblock {\em Physica D}, pages 1--39, 2003.

\end{thebibliography}

\end{document}